\documentclass[preprint,5p,times]{elsarticle}

\usepackage[T1]{fontenc}
\usepackage{amsmath}
\usepackage{amsthm}
\usepackage{breakurl}
\usepackage{breqn}
\usepackage[breaklinks]{hyperref}
\usepackage{amssymb}
\usepackage{lipsum}
\usepackage{listings}
\usepackage{gensymb}
\usepackage{fancyhdr}
\usepackage{graphicx}
\usepackage{url}
\usepackage{lineno}
\usepackage{xcolor,colortbl}
\usepackage{color}
\usepackage{multirow}
\usepackage{algorithm}
\usepackage{algpseudocode}

\definecolor{temporal}{RGB}{112, 186, 164}
\definecolor{red}{RGB}{219, 191, 186}
\definecolor{orange}{RGB}{191, 167, 214}

\journal{}

\begin{document}
\begin{frontmatter}

\title{Modeling GPU Dynamic Parallelism for Self Similar Density Workloads}

\author[a]{Felipe A. Quezada}
\author[a]{Crist\'obal A. Navarro\corref{author}}
\author[b]{Miguel Romero}
\author[c]{Cristhian Aguilera}
\cortext[author] {Corresponding author.\\\textit{E-mail address:} cnavarro@inf.uach.cl}
\address[a]{Instituto de Informática, Universidad Austral de Chile.}
\address[b]{Faculty of Engineering and Science of Universidad Adolfo Ibáñez.}
\address[c]{Universidad de los Lagos.}

\begin{abstract}
Dynamic Parallelism (DP) is a runtime feature of the GPU programming model that allows GPU threads to execute additional GPU kernels, recursively. Apart from making the programming of parallel hierarchical patterns easier, DP can also speedup problems that exhibit a heterogeneous data layout by focusing, through a subdivision process, the finite GPU resources on the sub-regions that exhibit more parallelism. However, doing an optimal subdivision process is not trivial, as there are different parameters that play an important role in the final performance of DP. Moreover, the current programming abstraction for DP also introduces an overhead that can penalize the final performance.   
In this work we present a subdivision cost model for problems that exhibit self similar density (SSD) workloads (such as fractals), in order understand what parameters provide the fastest subdivision approach. Also, we introduce a new subdivision implementation, named \textit{Adaptive Serial Kernels} (ASK), as a smaller overhead alternative to CUDA's Dynamic Parallelism.
Using the cost model on the Mandelbrot Set as a case study shows that the optimal scheme is to start with an initial subdivision between $g=[2,16]$, then keep subdividing in regions of $r=2,4$, and stop when regions reach a size of $B \sim 32$. The experimental results agree with the theoretical parameters, confirming the usability of the cost model. In terms of performance, the proposed ASK approach runs up to $\sim 60\%$ faster than Dynamic Parallelism in the Mandelbrot set, and up to $12\times$ faster than a basic exhaustive implementation, whereas DP is up to $7.5\times$.
These results put the subdivision cost model and the ASK approach as useful tools for analyzing the potential improvement factor from subdivision before-hand, and for implementing an efficient GPU-based subdivision approach when developing GPU libraries or fine-tuning specific scientific algorithms.
\end{abstract}


\end{frontmatter}

\section{Introduction}
In the last decade, GPU computing has become a widely available tool 
for the development of advanced and efficient HPC solutions for science and technology \cite{navarro2014survey, nickolls2010gpu}. The GPU programming model allows a programmer to describe scalable parallel algorithms that can accelerate data-parallel problems by up to an order of magnitude compared to a CPU-based solution \cite{owens2008gpu, bedorf2019bonsai, navarro2016adaptive}. From top to bottom, the programming model provides three constructs that allow mapping the work units to the data-domain. In the case of CUDA (Nvidia's GPU programming platform) these are named grid, block and thread \cite{nickolls2008scalable}. In the early age of general-purpose GPU computing (2007 to 2012), the programming model only allowed a static definition of these constructs, which made GPUs less friendly to deal with problems with a hierarchical structure or heterogeneous data layout.

In the year 2012, Nvidia introduced \textit{Dynamic Parallelism} (DP)\footnote{In 2013, Khronos Group also introduced Dynamic Parallelism to the specification of the OpenCL 2.0 standard \cite{kaeli2015heterogeneous}, contributing to the compatibility of DP beyond Nvidia GPUs.}  \cite{jones2012introduction}, a programming feature that allows GPU threads to launch new kernels recursively during GPU runtime, all this without the need to return to the CPU for synchronization. Dynamic Parallelism contributes in two ways; i) as a programmer-friendly abstraction to solve problems with a hierarchical structure, and ii) as a dynamic exploration method to solve problems with an heterogeneous work layout (\textit{i.e.}, some of its regions require more work, while other regions require less to none), focusing GPU resources where more parallelism is required. For problems that only benefit from i), the motivation to use DP is the easiness of expressing parallel hierarchical exploration with the hope of introducing the least overhead possible from the abstraction. Techniques such as sorting \cite{neelima2017kepler} fall into this type of problems. On the other hand, for problems that exhibit an heterogeneous work layout, the motivation to use DP is not only the easiness in programming a parallel hierarchy, but it is also the possibility to further increase the speedup of a GPU implementation given that parallelism can now be focused on the regions of actual work. Simulation of partial differential equations (PDE) \cite{abdelfattah2016performance} in sparse domains, Barnes-Hut n-body simulation \cite{bedorf2012sparse} and fractal generation/processing \cite{nogaj2003comparisons,NAVARRO2020158} are example applications that fall into this category, among others.
 
\begin{figure*}[ht!]
    \centering
    \includegraphics[scale=0.68]{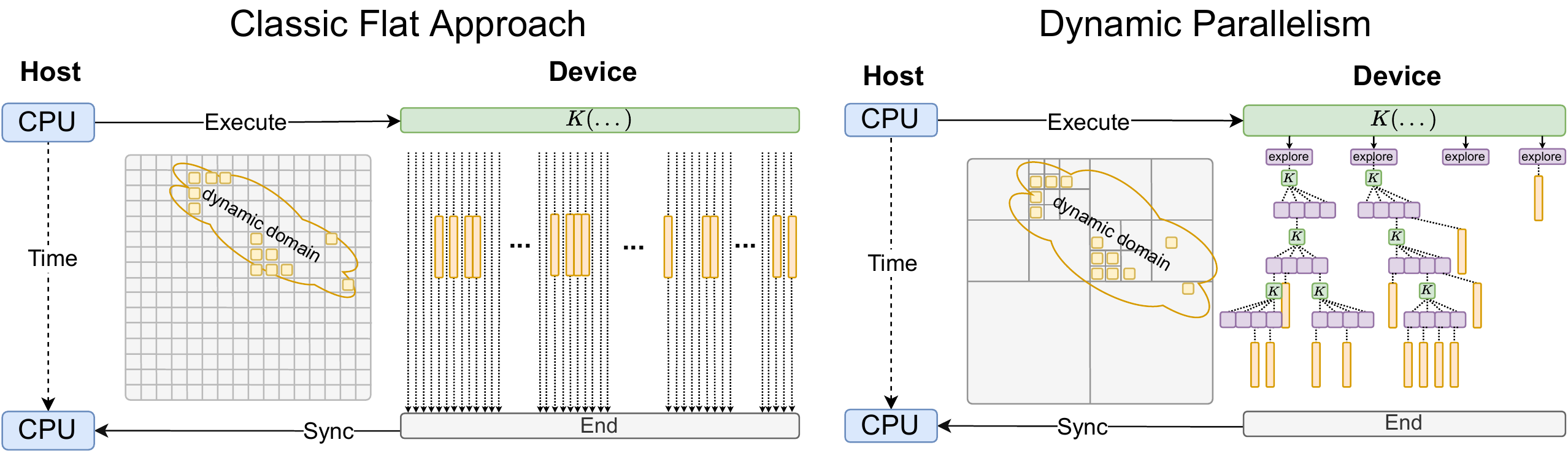}
    \caption{On the left, the classic GPU approach where many threads fall outside of the regions with work. On the right, DP first discovering the domain before executing parallel work.}
    \label{fig:dp-general}
\end{figure*}

DP has become a useful addition to the GPU programming model for doing a parallel subdivision process where parallel work discovered during execution. Figure \ref{fig:dp-general} (right) illustrates Dynamic Parallelism compared to a classic flat approach (left) when acting on a problem with a heterogeneous data layout.

Nowadays, DP is available on all modern GPUs through the CUDA and OpenCL platforms \cite{jones2012introduction, kaeli2015heterogeneous}, and has provided extra speedup in problems that exhibit an heterogeneous data layout \cite{10.1117/12.2018069, adinetz-dp, 7592731, jarzabek2017a}. For example, generating the Mandelbrot set is a well known favorable use case for DP \cite{mariani-silver, adinetz-dp}, where the subdivision based approach by DP reaches a significant extra speedup over a basic GPU implementation. There have also been several reports of performance slowdown from DP \cite{wang2014characterization, BOZORGMEHR2021104958, jarzabek2017a} because of the high overhead from recursive kernel executions along the subdivision process. In many cases this overhead can be reduced significantly if optimal values are used for three subdivision parameters; initial subdivision ($g$), recurrent subdivision ($r$) and minimal region size ($B$). Finding the optimal or near-optimal values for these $\{g,r,B\}$ parameters is not a trivial task as the configuration space is large and may vary on the problem size as well. Being able to analyze the efficiency of a DP-based subdivision process using a cost model would give key insights before hand for building a more efficient implementation of dynamic parallelism with optimal $g,r,B$ parameters and less overheads. Also, having a cost model motivates the proposal of new alternative subdivision methods that can introduce less overhead than the regular recursive DP.

This work provides two contributions: i) a cost model for the DP-based subdivision process, in which optimal parameters $g,r,B$ can be estimated for the Self Similar Density (SSD) class of problems, and ii) \textit{Adaptive Serial Kernels} (ASK), an alternative subdivision approach that executes a series of kernels, one after another, adapting the compute grids with the help of an offset lookup table that is tight in memory usage. Using the Mandelbrot Set as a case study, the proposed cost model was able to estimate the optimal $g,r,B$ values which were also confirmed by experimentation. Experimental results also showed that the proposed ASK subdivision approach is up to $\sim 12\times$ faster than an exhaustive approach, whereas the regular CUDA DP is up to $7.5\times$ faster. This translates to ASK being up to $\sim 60\%$ faster than DP. These results can be of great interest to GPU developers who seek to fine tune GPU-based library, a low level GPU algorithm, or even consider to improve the proposed cost model and alternative approach further beyond. 

The rest of the manuscript is organized as follows: Section \ref{sec:related-work} describes related work on the subject of GPU Dynamic Parallelism, Section \ref{sec:dynamic-parallelism} revisits GPU Dynamic Parallelism and explains how it is frequently utilized, Section \ref{sec:work-model} presents the work model in order to better understand the relevant parameters and the cost of a subdivision approach. Section \ref{sec:ask-formulation} proposes the alternative subdivision approach \textit{Adaptive Serial Kernels} and Section \ref{sec:extending-3D} describes the main aspects to consider when extending both regular DP and ASK to higher dimensions. Section \ref{sec:experimental-results} presents experimental performance results using the Mandelbrot Set as a case study and Section \ref{sec:conclusions} discusses and concludes the main results of this work.

\section{Related Work}
\label{sec:related-work}
Related work on GPU Dynamic Parallelism (DP) can be classified into two groups, i) parallel algorithms developed with DP and ii) analysis/improvements to the DP model. In general, DP's recursion overhead is recognized as a critical aspect that affects performance.

\subsection{DP-based Parallel Algorithms}
DiMarco and Taufer (2013) \cite{10.1117/12.2018069} developed parallel DP algorithms for k-means and hierarchical clustering. The authors report speedups of up to $\sim3\times$ for hierarchical clustering and a slowdown of $7.7\%$ for k-means. 
Zhang et. al. (2015) \cite{10.1145/2833179.2833189} proposed DP versions of breath first search and single source shortest paths. The authors report that the performance was not as efficient as expected, but on the other hand compensates in the ease of programming.
Alandoli \textit{et al.} \cite{7592731} used DP for accelerating cluster-based community detection in social networks. The authors reported speedups of up to $4.45\times$ over a sequential implementation, but it was not as fast as an hybrid CPU-GPU approach.
In 2021, Bozorgmehr\textit{et al.}~\cite{BOZORGMEHR2021104958} reported competitive performance of DP when simulating 3D wind-field phenomena, but it was not as fast as an hybrid CPU-GPU implementation following classic strategies.

\subsection{Analysis/Improvements to DP}
In 2014, Wang and Yalamanchili~\cite{wang2014characterization} analyzed the mayor causes of overhead in DP, finding that these are the recursive calls to child kernels and a low cache hit-ratio for dynamically allocated memory. 
The authors also report an average slowdown of $1.21\times$ for algorithms like BFS, Graph coloring, regular expression Match, relational join, among others, and identify that if no execution overhead existed, DP could have provided a speedup as high as $2.73\times$.

In 2016, Plauth et al.~\cite{MaxPlauth2016} studied the overhead of DP using the N-Queen problem as case study. Four different incremental DP approaches were proposed to solve the problem. The authors conclude that in all variants 
the performance benefits were outweighed by the overhead from nested child kernel launches and dynamic memory allocation. Other studies have focused in designing mechanisms that help the programmer implement dynamic parallelism. Such is the case of Li, Wu and Becchi~\cite{10.1145/2723772.2723780,7516050}, whom in 2015 proposed a workload consolidation mechanism for irregular nested loops using dynamic parallelism. 
Their main results are speedups from $90\times$ to $3300\times$ over a basic DP approach, and from $2\times$ to $6\times$ over a flat GPU implementation. 

In 2017, Jarz{\k{a}}bek and Czarnul~\cite{jarzabek2017a} analyzed the performance of Dynamic Parallelism and Unified Memory under three different applications: heat transfer simulation in 2D space, adaptive integration with a trapezoidal rule and a verification of the Goldbatch conjecture. Results show mixed performances: for the heat transfer simulation, speedups were favorable. For adaptive integration, DP performed similar but slightly worse compared to the standard approach. For the Goldbatch conjecture, DP under-performed considerably. Authors claim that incorporating DP in code was not trivial and increased code complexity. Finally they conclude that DP can bring considerable benefits for recursive algorithms or algorithms that use hierarchical arranged data. 

In 2017, Tang \textit{et. al.}~\cite{7920863} proposed SPAWN, a runtime framework that controls the launch of child kernels, reducing overhead associated with kernel launches and queue latency. Additionally, the work allows a better mix of child and parent kernels for the scheduler to effectively hide the remaining overheads and improve the utilization of the GPU resources. Their results show that using SPAWN provides from $57\%$ to $69\%$ of speedup over DP and flat GPU implementations, respectively.
In 2016, El Hajj \textit{et. al.}~\cite{7783716,el2018techniques} proposed a set of compile-time techniques that reduce the total number of child kernel spawned and increase the amount of work done per kernel. This is done by aggregating kernels in the same warp, block, or kernel. Results show that kernel aggregation achieves up to $6.58\times$ speedup over regular DP.

In general, recent research on DP has recognized the performance overhead of recursive GPU kernels, which limits the potential performance of a subdivision based approach. Even when in some cases DP does provide an extra speedup over a classic exhaustive GPU approach, this speedup could be higher if the subdivision parameters $\{g,r,B\}$ are also considered in the analysis. The next Section revisits the main features of GPU Dynamic Parallelism.


\section{Revisiting GPU Dynamic Parallelism}
\label{sec:dynamic-parallelism}
Dynamic Parallelism (DP) is a GPU programming abstraction that allows GPU threads to launch additional GPU kernels within a given kernel and produce nested parallelism at execution time. This feature facilitates the programming on problems that exhibit a hierarchical structure, and can potentially produce an extra speedup for problems with an heterogeneous data layout, as it allows focusing the GPU resources (threads) on the regions that exhibit parallel work while avoiding the use of resources in regions with little or no parallelism. Figure \ref{fig:dp-general} (right) illustrates how extra speedup can potentially emerge by focusing parallel resources. 

In terms of programming, the DP approach offers a tree-like structure of recursive kernel calls, traversed top-down. The root kernel is launched from the host as a regular CUDA kernel, while the rest of the kernels are called recursively from their parent kernels. Child kernels are recursively generated as long as more parallel work is found at runtime for that region of the data-parallel domain. 
%
Internal nodes of the execution tree play the role of \textit{exploration kernels}, \textit{i.e}, kernels that besides some amount of application work done, are also in charge of discovering if further parallelism is required or not. On the other hand, leaf nodes play the role of \textit{work kernels}, \textit{i.e.}, kernels that focus on doing the original parallel-work intended for the algorithm.

\begin{figure}[ht!]
    \centering
    \includegraphics[scale=1.0]{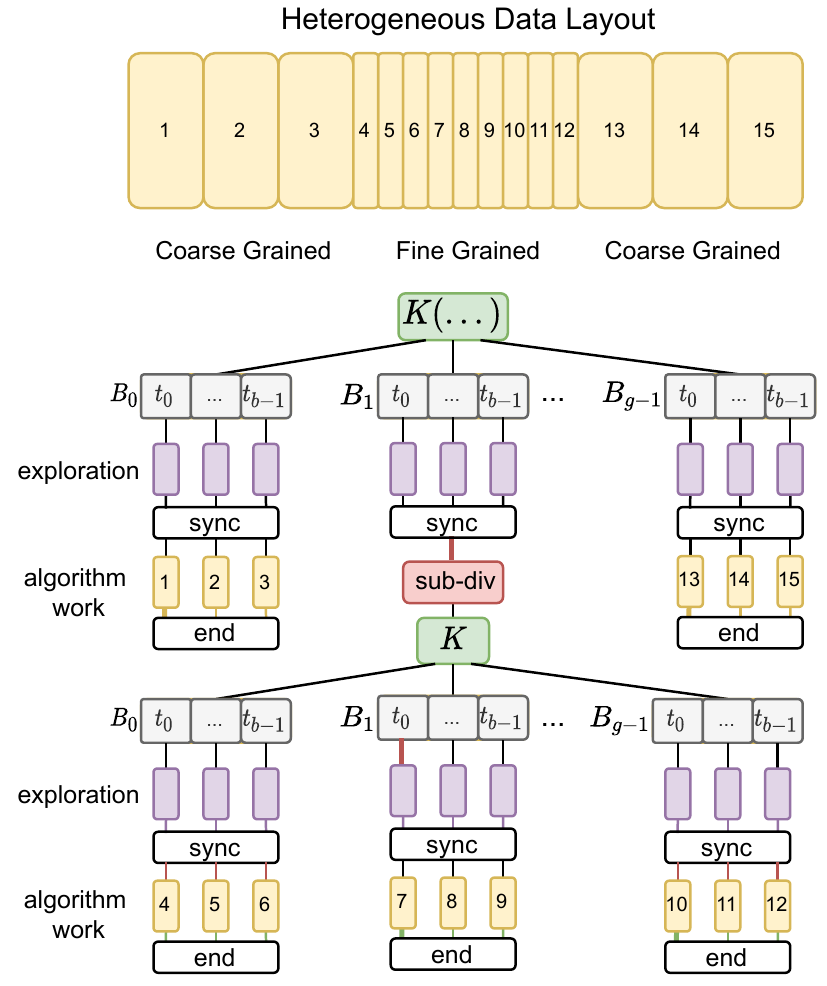}
    \caption{Dynamic Parallelism acting on an heterogeneous data-parallel problem domain.}
    \label{fig:dp-detailed}
\end{figure}

Figure \ref{fig:dp-detailed} illustrates DP's execution tree on a heterogeneous data layout, where the middle block subdivides and launches a new parallel kernel in order to handle the finer grained paralleism exhibited by the problem. 

It is worth noticing the red sub-div block of Figure \ref{fig:dp-detailed}, which illustrates the existence of a subdivision cost in the process. More details about DP can be found in Jones's presentation on Dynamic Parallelism \cite{jones2012introduction}. 

Although Dynamic Parallelism is a tool that facilitates expressing nested parallelism for hierarchical structures, it has also been used to obtain extra speedup for problems that exhibit a heterogeneous data layout given that resources can be dynamically mapped onto the regions that require more work. This potential speedup may not be achievable if the subdivision parameters are too distant from its optimal values, as it may introduce too much kernel recursion overhead. Having a cost model that can measure the cost of a subdivision-based approach can provide useful insights in order to build an efficient subdivision process with optimal parameters and reduced overhead. The next Section formulates a cost model in order to understand what are the parameters that play a relevant role in a subdivision process, and find which values contribute to an efficient solution.

\section{Cost Model for Hierarchical Subdivisions}
\label{sec:work-model}
This Section analyses the work and parallel time for both the exhaustive parallel approach and the subdivision-based approach. The analysis focuses on problems that exhibit self similar density which is a type of heterogeneous data layout. For these problems, a subdivision based approach can provide extra speedup over an exhaustive approach. This theoretical analysis leads to the formulation of a cost model that contributes in the understanding of what parameters play a relevant role on a subdivision-based process, and what combination of values produce optimal performance. 

\subsection{Exhaustive Work}
The exhaustive approach maps resources to all data locations in a flat GPU kernel. The computation uses a fixed-size grid $G_E$ that maps threads to data elements one-to-one, for the whole $n\times n$ domain. The grid is then processed with an application work of $\mathcal{A}$ per data element, giving a total work of
\begin{align}
    \label{eq:exhaustive-general}
    W_E(n) &= |G_E|\mathcal{A}\\
           &= n^2 \mathcal{A}.
\end{align}
The value of $\mathcal{A}$ corresponds to the application's algorithm work. It is normally assumed that the domain exhibits sufficient data-parallel elements to produce a potential speedup. 

\subsection{Sub-division Work}
For a subdivision approach, its total work denoted $W_S(n)$ is the sum of the work found at all subdivision levels, denoted $K(n,\tau)$, added with the cost of doing algorithm work in the last level, denoted $L(n,\tau)$, \textit{i.e.}, 
\begin{align}
    \label{eq:work-subdivision-general}
    W_S(n) = K(n, \tau) + L(n,\tau).
\end{align}
In Eq.~(\ref{eq:work-subdivision-general}), symbol $\tau$ is the number of sub-division levels, equivalent to the depth, and it is normally defined as a function of $n$. Expanding $K(n,\tau)$ we have:
\begin{align}
    \label{eq:work-subdivision-K}
    K(n,\tau) = \sum_{i=0}^{\tau-2} K_i(n) 
\end{align}
where $K_i(n)$ is the work at the $i$-th level of the subdivision process. Given that the summation is of the type zero-index (\textit{i.e.}, last level is $\tau -1$), and that the last level is treated separately, the summation goes only up to $\tau-2$. In practice, the actual value of $\tau$ in a specific application depends on the grid resolution, the subdivision factor and the dynamics of the application problem as well, meaning that even under equal grid and subdivision conditions, two problems with different dynamics may still produce a different value of $\tau$. Because of all the possible ways an application can behave across its domain, we represent this aspect as if, at a certain depth level, subdivision occurs with a certain probability, meaning that regions of the $i$-th level subdivide with probability $P_i \in [0,1]$, and stop subdividing with probability $(1-P_i)$. In this representation, the work of the $i$-th level can be expressed as
\begin{align}
    K_i(n) &= \sum_{\rho = 1}^{\mathbb{E}[|G_i|]} \bigg({\mathbb{E}[W_S^{i,\rho}] + \mathbb{E}[W_T^{i,\rho}]}\bigg)
    \label{eq:work-sub-1}
\end{align}
where the cost of a given level is the sum over the work of its expected number of regions, \textit{i.e.}, $\mathbb{E}[|G_i|]$, considering sub-division ($W_S^{i,\rho}$) and algorithm ($W_T^{i,\rho}$) work. It is worth noticing that under this setting, at a given depth level all regions have the same probability of subdividing, therefore the expectation values for $\mathbb{E}[W_S^{i,\rho}]$ and $\mathbb{E}[W_T^{i,\rho}]$ are the same for each $\rho \in \mathbb{E}[|G_i|]$ and thus can be written simply as $\mathbb{E}[W_S^i]$ and $\mathbb{E}[W_A^i]$, respectively. 
With this simplification, the expectation of the subdivision work for a region at the $i$-th level, \textit{i.e.}, $W_S^i$, is 
\begin{align}
    \mathbb{E}[W_S^i] = P_i (Q + S)
\end{align}
with $Q$ being the cost of an exploration query in order to know if a region requires further subdivision or not, and $S$ being the cost of subdividing that region. The expectation value of the application's terminal work (\textit{i.e.}, ending a subdivision branch), denoted $W_T^i$, becomes 
\begin{align}
\mathbb{E}[W_T^i] = (1-P_i)(Q + T)
\end{align}
with $T$ being the work involved for discarding the region and terminating the region with a final work pass. Using these definitions in Eq.~(\ref{eq:work-sub-1}) leads to
\begin{align}
    K_i(n) &= \mathbb{E}[|G_i|] \bigg({P_i(Q + S) + (1-P_i)(Q + T)}\bigg).
\end{align}

For $\mathbb{E}[|G_i|]$, its value depends on the expectation of its previous grid size, recursively. Given that $i-1$ levels precede $i$, $\mathbb{E}[|G_i|]$ can be expressed as
\begin{align}
    \label{eq:expectation-grid-recursive}
    \mathbb{E}[|G_i|] &= P_{i-1} R \mathbb{E}[|G_{i-1}|],\ \ i  = 1..\tau-2\\
    \mathbb{E}[|G_0|] &= G
\end{align}
where $R = r_x \times r_y$ is a chosen subdivision and $G = g_x \times g_y$ is the initial number of regions, or starting subdivision, which can be different from $R$ in order to begin with a more populated grid and favor parallelism at an early stage. The iterative form of the recurrence from Eq. (\ref{eq:expectation-grid-recursive}) is
\begin{equation}
    \label{eq:expectation-grid-iterative}
    \mathbb{E}[|G_i|] = G R^{i} \prod_{j=0}^{i-1} P_j.
\end{equation}

Defining the auxiliary term $U_i = P_i(Q + S) + (1-P_i)(Q + T)$, we obtain the following work expression for $K_i(n)$,
\begin{align}
    \label{eq:work-subdiv-general}
    K_i(n)  &= U_i G R^{i} \prod_{j=0}^{i-1} P_j.
\end{align}

For the last-level work, \textit{i.e.}, $L(n,\tau)$ from Eq.~(\ref{eq:work-subdivision-general}), it considers the algorithmic work performed on all data-elements that remain after the last subdivision of $K(n,\tau)$. Its expression is defined by considering the size of regions at the last level, $(\frac{n^2}{G R^{\tau-1}})$, multiplied by the algorithmic cost per-thread and by the expected number of sub-regions that would exist given the accumulated probability up to that level, \textit{i.e.},
\begin{align}
    L(n,\tau) &= \bigg( \frac{n^2}{G R^{\tau-1}} \bigg)\mathcal{A} G R^{\tau - 1} \prod_{j=0}^{\tau-2} P_j.\\
    \label{eq:work-subdiv-last-level}
              &= n^2 \mathcal{A} \prod_{j=0}^{\tau-2} P_j.
\end{align}
Using the two formulations from Eqs. (\ref{eq:work-subdiv-general}) and (\ref{eq:work-subdiv-last-level}), the total subdivision cost $W_S(n)$ becomes
\begin{align}
    \label{eq:work-subdiv-final-exp-general}
    W_S(n) &= K(n,\tau) + L(n,\tau)\\
    \label{eq:work-subdiv-final-exp-general-expanded}
         &= \left[ \sum_{i=0}^{\tau-2}\bigg(U_i G R^{i} \prod_{j=0}^{i-1} P_j\bigg)\right] + n^2 \mathcal{A} \prod_{j=0}^{\tau-2} P_j.
\end{align}
Eq.~(\ref{eq:work-subdiv-final-exp-general-expanded}) represents a general work expression for problems that can exhibit different subdivision probabilities at each depth level. If the set of probabilities $P_0,P_1,\dots, P_{\tau-2}$, and the parameters $Q, T, \mathcal{A}, S$ are well defined for a certain application, then Eq.~(\ref{eq:work-subdiv-final-exp-general-expanded}) can be used to estimate the computational work required by a subdivision approach, and assist in finding proper values of $G,R,\tau$ for an optimal work scheme, or one that is sufficiently efficient for the application.

\subsubsection{Subdivision Cost on the SSD Sub-class}
A sub-class of subdivision-friendly problems can be defined by introducing the following three assumptions:
\begin{enumerate}
    \item[i)] The subdivision probability remains equal or highly similar through its levels, \textit{i.e}, $P_0 \sim P_1 \sim ... \sim P_{\tau-1}$.
    \item[i)] Subdivisions are regular $\rightarrow$ $r = r_x = r_y$, and $g = g_x = g_y$.
    \item[iii)] The subdivision depth limit is $\tau = \log_{r}(\frac{n}{gB})$, where $B$ is an arbitrary block-size value.
\end{enumerate}
The first assumption produces a simpler sub-class of problems where work density has a self-similar property across its depth levels, that is, the work density found at any region, at any sub-division level, is similar to the work density found in the whole problem domain. We will refer to this class of problems as Self-Similar-Density (SSD) problems. An example type of application that belongs to the SSD class are fractals (among others), which due to their geometric self-similarity property, exhibit Self-Similar-Density as well. 

Assumption i) simplifies $K_i(n,\tau)$ into $\prod_{j=0}^{i-1} P_j = P^{i}$ as well as  $L(n,\tau)$ into $\prod_{j=0}^{\tau-2} P_j = P^{\tau-1}$. These changes lead to a work expression of 
\begin{align}
    \label{eq:work-subdiv-final-exp-self-similar}
    W_{\mathcal{SSD}} &= \sum_{i=0}^{\tau-2}K_i(n, \tau) + L(n,\tau)\\
    \label{eq:work-subdiv-final-exp-self-similar-QT-simplified}
         &= \left[ \sum_{i=0}^{\tau-2} \bigg(Q + PS + (1-P)T\bigg) G R^{i} P^{i} \right] + n^2\mathcal{A} P^{\tau-1}.
\end{align}
Assumption ii) makes the analysis easier without loss of generality, while assumption iii) allows the subdivision process to stop at an early stage if wanted, which turns out to be important for the remaining analysis as there can exist non-trivial optimal values for the stopping size.

In order to further analyze $W_{\mathcal{SSD}}(n)$, the parameters $Q$, $T$, $\mathcal{A}$ and $S$ need to be specified in the context of an application. In this work, we choose the Mandelbrot Set as a case study and the Mariani-Silver algorithm \cite{mariani-silver} as the process for generating it through subdivision. Also, the Mandelbrot Set is a known case of dynamic-parallelism being applied successfully with significant speedup. In the Mariani-Silver algorithm, the cost for the exploration query is $Q = \frac{4n\mathcal{A}}{gr^{i}}$ as it computes the \textit{dwell} (algorithm work) on the perimeter of a sub-region at the $i$-th depth level. If the perimeter pixels end with equal dwell values, then the subdivision process stops and a terminal work $T$ is applied to the region. In this case, the terminal work is $T = \frac{n^2}{G R^{i}}$ as it writes a constant value on each data-element. If the perimeter is heterogeneous, then the algorithm subdivides the region into $r_x \times r_y$ sub-regions with a sub-division cost of $S$. The value of $S$ does depend on the sub-division approach, that is, if using a recursive-based approach such as in DP, or an iterative one. Given that $S$ also relates to the GPU hardware and its latency produced by the subdivision approach, we opted to express $S$ relative to the application's algorithm cost, \textit{i.e.}, $S = \lambda \mathcal{A}$. In the case of the Mandelbrot Set, $\mathcal{A}$ is often chosen arbitrarily and corresponds to the dwell which is the number passes done on the dynamical system equation, where more passes lead to higher precision on deciding convergence/divergence. 

Updating $K_i(n, \tau)$ from Eq. (\ref{eq:work-subdiv-final-exp-self-similar-QT-simplified}) we get
\begin{align}
    K_i(n,\tau)  &= \bigg[\frac{4n\mathcal{A}}{g r^{i}} + P(\lambda \mathcal{A}) + (1-P)\frac{n^2}{G R^{i}}\bigg] G R^{i}P^i.
\end{align}
With this, the total cost of the subdivision approach applied for the Mandelbrot case, denoted $W_{\mathcal{SSD}}^{\mathcal{M}}$, now becomes
\begin{dmath}
          \label{eq:work-subdiv-final-exp-self-similar-expanded}
          W_{\mathcal{SSD}}^{\mathcal{M}} = \left[\sum_{i=0}^{\tau-2} \bigg[\frac{4n\mathcal{A}}{g r^{i}} + P(\lambda \mathcal{A}) + (1-P)\frac{n^2}{G R^{i}}\bigg] G R^{i}P^i\right] + n^2 \mathcal{A} P^{\tau-1}. 
\end{dmath}

\subsubsection{Work Reduction Factor of $W_{\mathcal{SSD}}^{\mathcal{M}}$}
The theoretical work improvement produced by a subdivision approach over an exhaustive one will be referred to as the \textit{work reduction factor}, which is the quotient between the Exhaustive and Subdivision work quantities. 
In the case of Mandelbrot Set, the work-reduction-factor is
\begin{align}
    \Omega = \frac{W_E}{W_{SSD}^{\mathcal{M}}}
\end{align}

\begin{figure*}[ht!]
    \centering
    
    \includegraphics[scale=0.59]{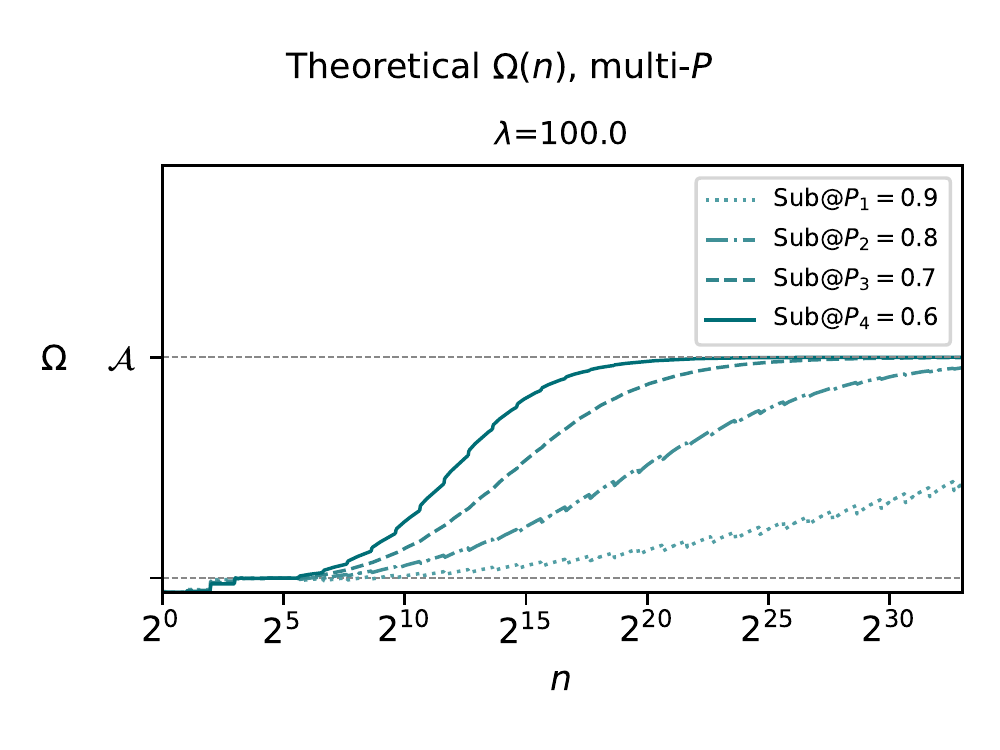}
    \includegraphics[scale=0.59]{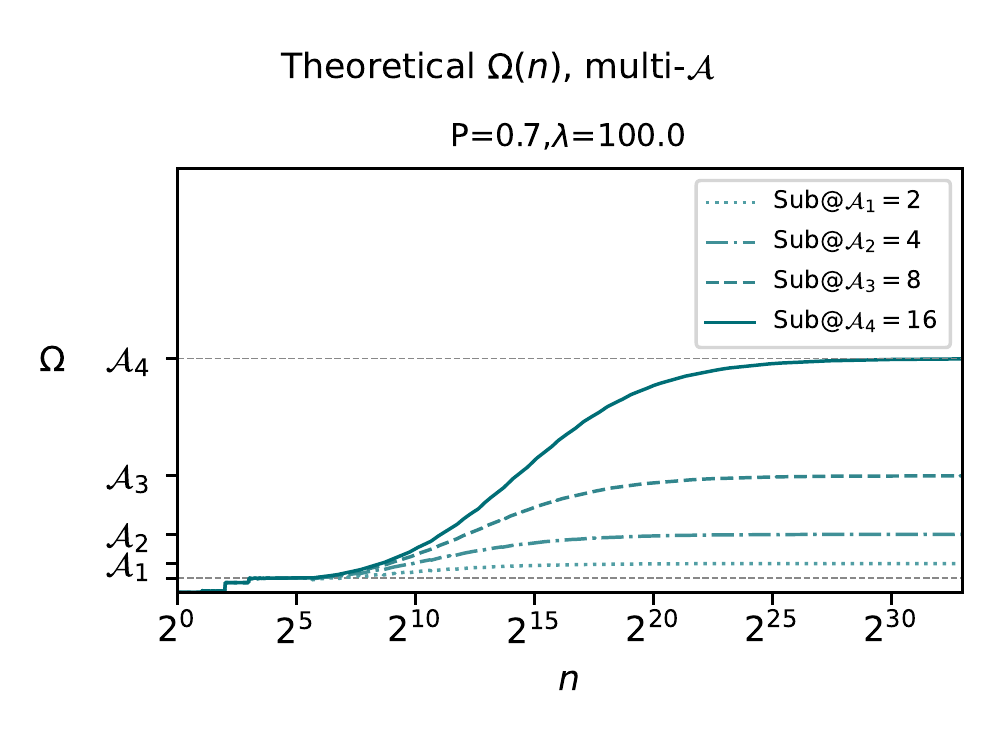}
    \includegraphics[scale=0.59]{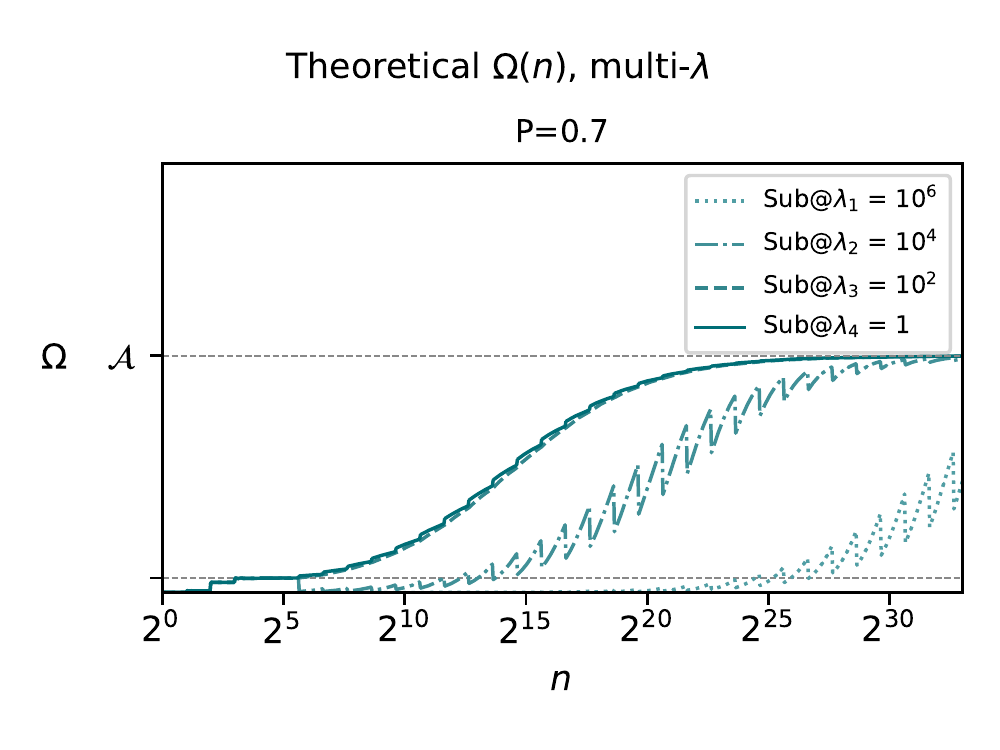}\\
    
    \includegraphics[scale=0.59]{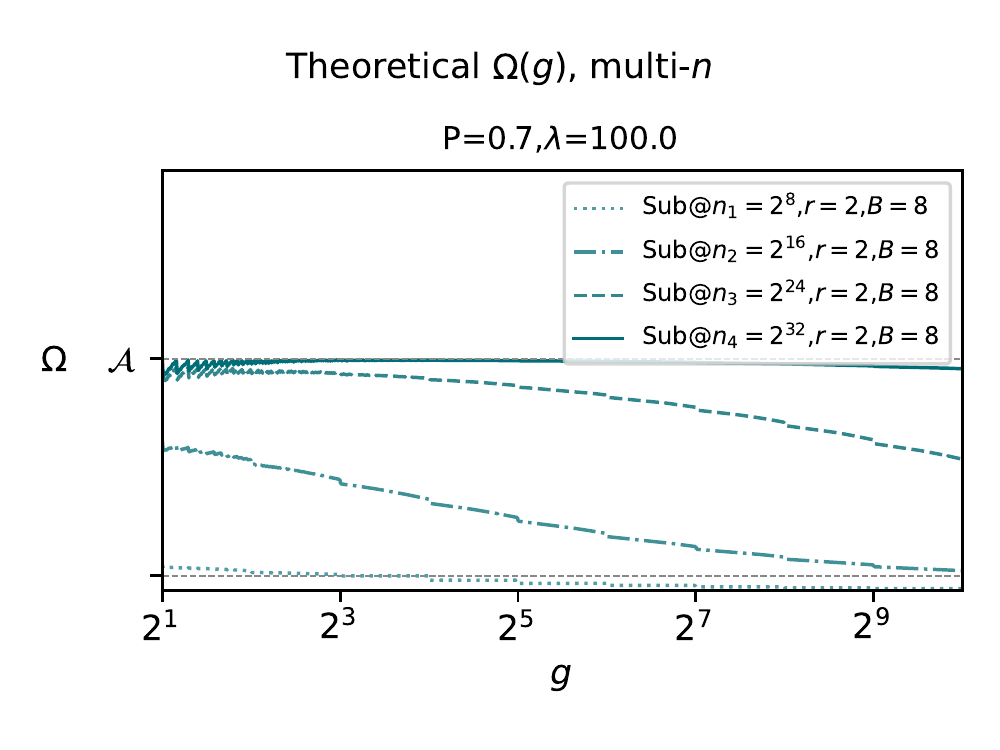}
    \includegraphics[scale=0.59]{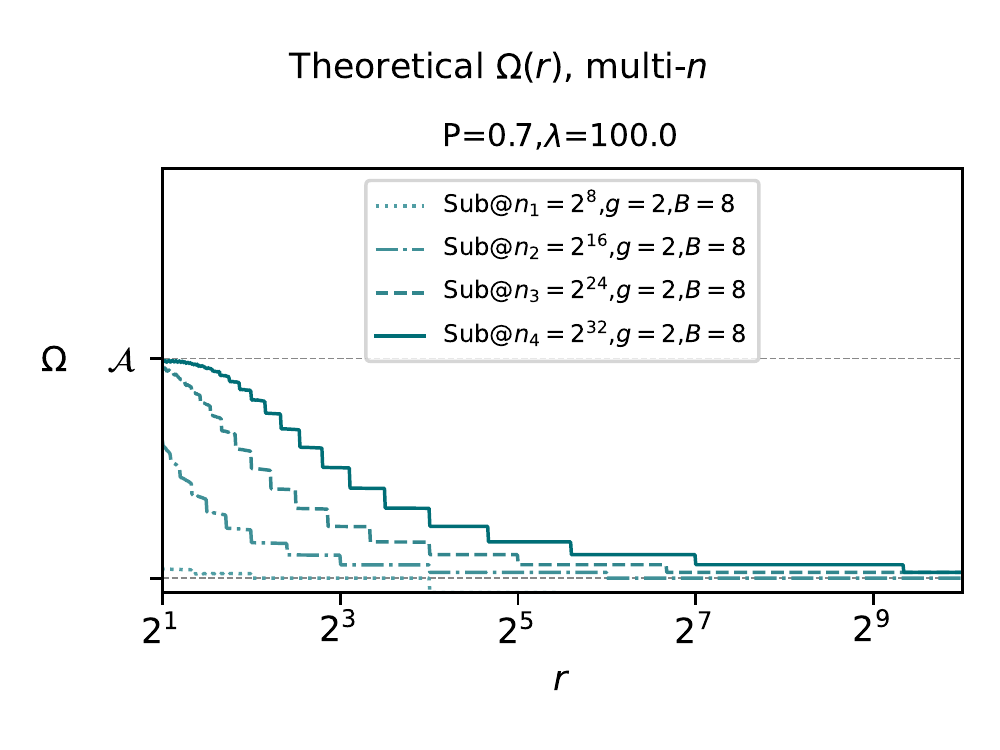}
    \includegraphics[scale=0.59]{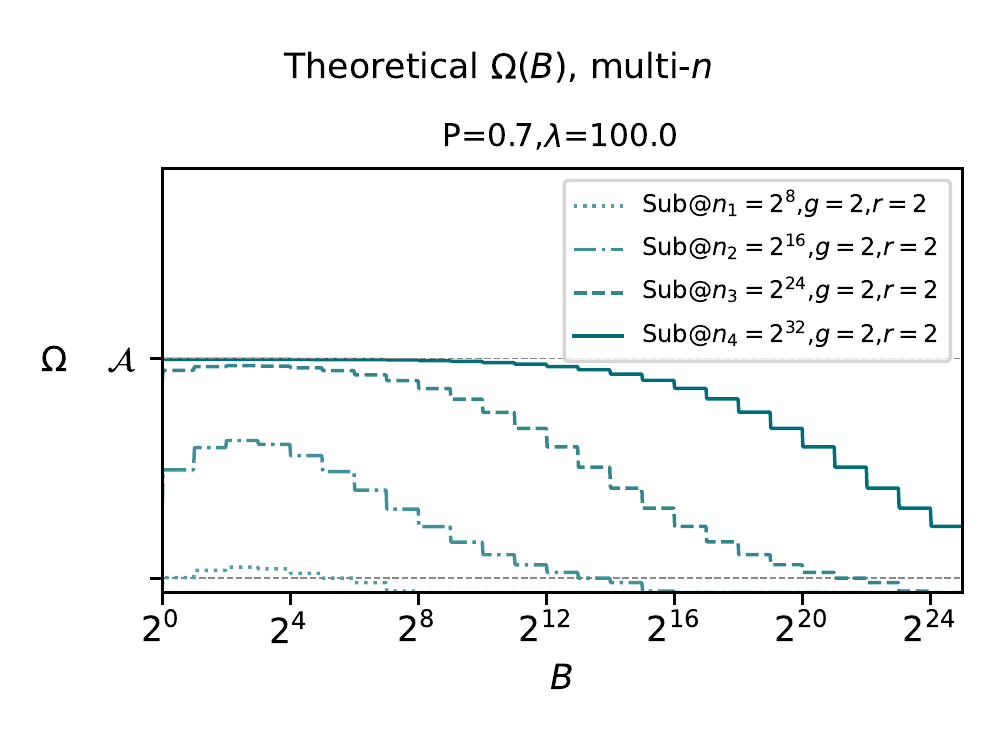}\\
    
    \includegraphics[scale=0.59]{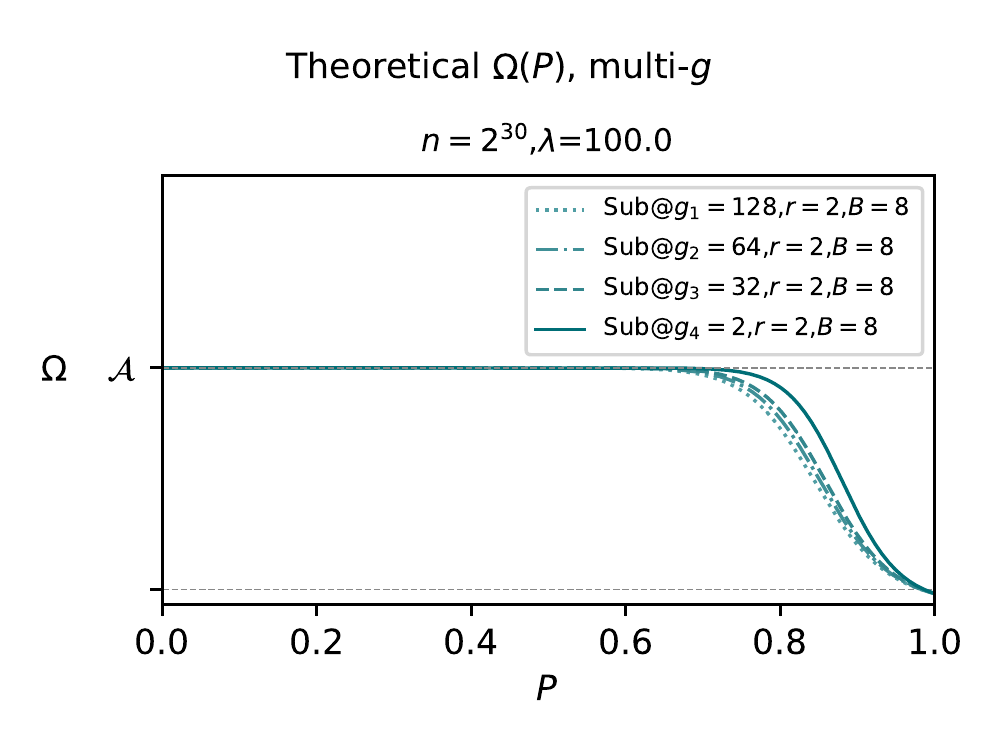}
    \includegraphics[scale=0.59]{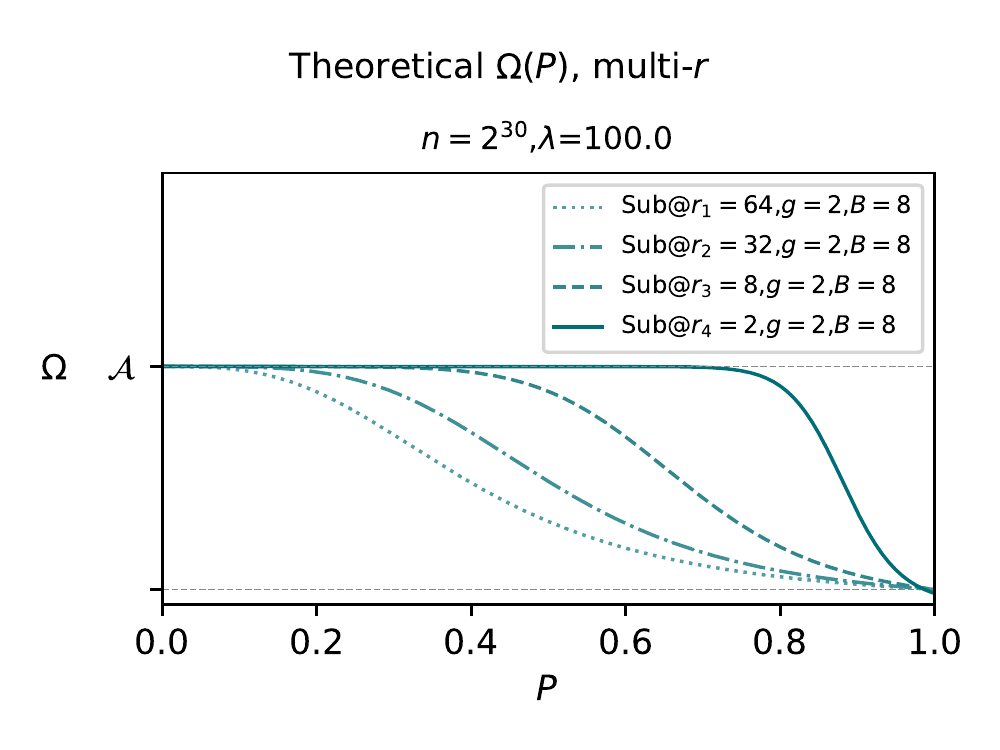}
    \includegraphics[scale=0.59]{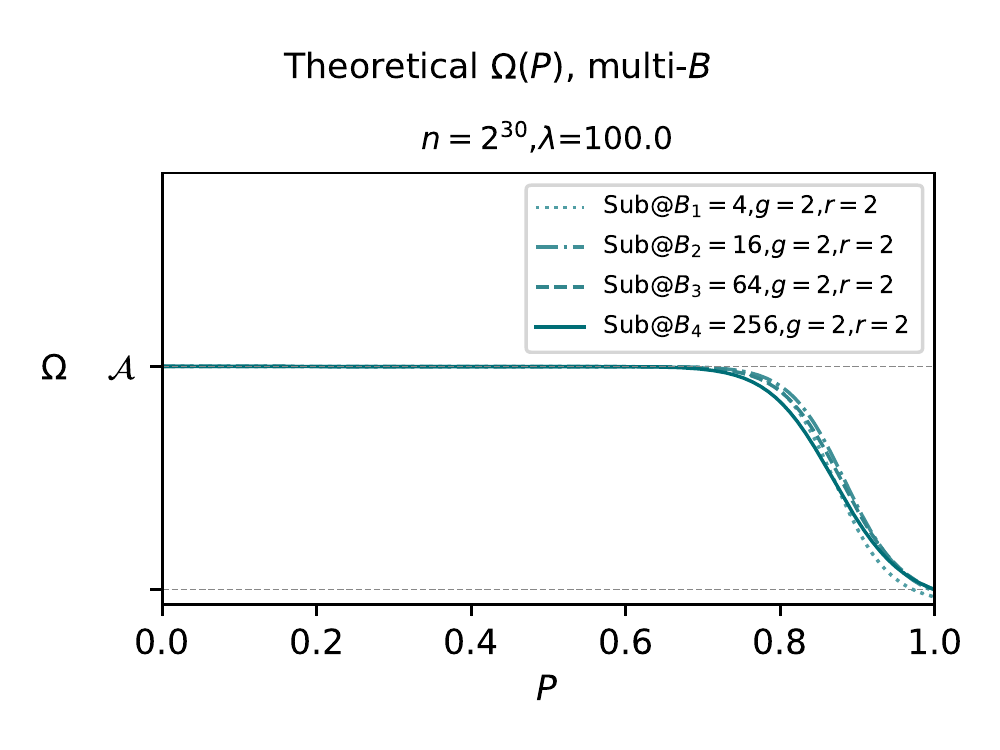}
    
    \caption{$\Omega$: Work-reduction-factor  (Eq.~(\ref{eq:work-subdiv-final-exp-self-similar-expanded})) plotted varying different parameters ($n, g, B, r, P$).}
    \label{fig:wrf-analytic}
\end{figure*}

Figure \ref{fig:wrf-analytic} presents a set of plots for 
$\Omega$, exploring the effect of different parameters, always choosing the optimal values of $g,r,B$ (\textit{i.e}, the $r,g,B$ values that lead to the minimum work for the subdivision approach) in a search space of $\{2,4,8,\cdots,1024\}$ for each one. 
For all plots, one can notice that the maximum speedup is upper bounded by $\mathcal{A}$. 

The first row of plots show the behavior of $\Omega$ as a function of $n$, varying $P$, $\mathcal{A}$ and $\lambda$, respectively. Lower probabilities make $\Omega$ reach its maximum work reduction factor value sooner in $n$ and diminishes as it approaches $P \sim 1$. In the large-$n$ regime, all curves eventually meet at the upper bound of $\mathcal{A}$. The third plot shows how the cost of subdivision affects $\Omega$ more in the medium to low $n$ regime, and less in the large-$n$ regime. 

The second row of plots shows in detail what are the optimal values for $g,r,B$ at different values of $n$. In the case of $g$, smaller problem sizes require a smaller value, while the largest problem size suggests an optimal $g \sim 2^{4}$. For $r$, a value of $r \sim 2$ is the optimal. In the case of $B$, less work is done near $B \sim 2^3$.

The third row of plots present $\Omega$ as a function of $P$, using four different values for $g$, $r$ and $B$. In the case of the multi-$g$ plot, there is a transition point at $P \sim 0.7$ in which $\Omega$ starts to reduce. For the multi-$r$ plot, the transition point varies depending of the value of $r$. In the case of multi-$B$, it has a lesser impact on the transition point.

These theoretical work results show that the highest value of $\Omega$ is achieved when using a subdivision stopper of $B \sim 2^3$, with subdivision values of $r \sim 2$ and initial subdivision of $g \sim 2^1$ for small problems and $g \sim 2^4$ for large problems. Converting work to parallel time can modify these optimal values, as it requires applying a parallelization scheme.

\subsection{Theoretical Time and Speedup}
A two-level GPU computing model is employed on the work expressions in order to obtain parallel time and speedup. At the top level there are $q$ multiprocessors that have a PRAM-like access to memory (a global shared memory), but no\footnote{Unless special features are used, such as CUDA's cooperative groups.} synchronization during a kernel execution. At the bottom level, there are $c$ cores per multiprocessor, with a PRAM-like behavior that includes synchronization during kernel execution. Using this model, the core count of a GPU becomes $q \times c$. 
For the Exhaustive approach, a flat-parallel GPU scheme produces a running time of 
\begin{equation}
    \label{eq:time-exhaustive} 
    T_{Ex}^{\mathcal{M}} = \left \lceil \frac{n^2}{qc} \right\rceil \mathcal{A}.
\end{equation}
This scheme is already efficient and locality can be exploited at each block whenever is possible.
For the subdivision-based approach there are two parallel schemes that can be applied:
\begin{itemize}
    \item[i)] SBR: Single-Block per region.
    \item[ii)] MBR: Multiple-Blocks per region.
\end{itemize}
The difference between SBR and MBR, \textit{i.e.}, single vs multiple blocks per region, has an impact on how GPU resources are mapped to the problem domain, how synchronized they are, and puts into manifest a trade-off between synchronization (SBR) and core quantity (MBR). In the SBR scheme each region is processed by only one block of $c$ synchronized cores, whereas in MBR each region is handled by multiple-blocks of $c$ cores, but these blocks of cores cannot synchronize easily among themselves. 

\subsubsection{Parallel Scheme i): Single-Block per Region (SBR)}
The SBR parallelization scheme works by processing each region by one of the $q$ multiprocessors, leaving $c$ cores available to do the work for $Q_i$ or $T_i$. The application of SBR parallelism is denoted as $\Delta[x]$, with $x$ being a given work expression. Applying SBR into $W_{\mathcal{SSD}}^{\mathcal{M}}$ from Eq. (\ref{eq:work-subdiv-final-exp-self-similar-QT-simplified}) produces   
\begin{dmath}
    \label{eq:time-subdiv-QT}
     T_{\textit{SBR}}^{\mathcal{M}} = \sum_{i=0}^{\tau-2} \bigg(\Delta[Q_i] + PS + \Delta[(1-P)T_i]\bigg) \Delta[G (RP)^i] + \Delta[L(n,\tau)]\\
    = \sum_{i=0}^{\tau-2} \left(\left \lceil \frac{4n}{g r^{i} c} \right \rceil \mathcal{A} + P\lambda \mathcal{A} + (1-P)\left \lceil \frac{n^2}{G R^{i}c} \right \rceil \right) \left \lceil \frac{GR^i}{q}\right \rceil P^i + \mathcal{A} \left \lceil \frac{n^2}{G R^{\tau-1} c} \right \rceil \left \lceil \frac{G R^{\tau - 1}}{q} \right \rceil P^{\tau-1}.
\end{dmath}
In the Mandelbrot Set, the query $Q_i$ is a parallel computation at the border pixels of each region. In the case of $T_i$, its work is a parallel write on the whole region. The term $L(n,\tau)$ is used as in Eq. (\ref{eq:work-subdiv-last-level}) before the region terms become simplified, this way the SBR scheme gets applied properly for $q$ and $c$. The ceiling functions are required to keep a minimum of $1$ unit of time. 

\subsubsection{Parallel Scheme ii): Multiple-Blocks per Region (MBR)}
The MBR scheme works by mapping multiple blocks of $c$ cores to each region if the number of resources allow it. In the case of the Mandelbrot Set, $T_i$ and $L$ provide sufficient parallel work to apply MBR, while $Q_i$ (boundary work) and $PS$ (subdivision cost) are less parallel, therefore these last two terms still use the SBR approach. The application of the MBR parallel scheme will be denoted as $\nabla[x]$, with $x$ being a given work expression. Applying it into $W_{\mathcal{SSD}}^{\mathcal{M}}$ from Eq. (\ref{eq:work-subdiv-final-exp-self-similar-QT-simplified}) produces
\begin{dmath}
    \label{eq:time-subdiv-QT}
         T_{\textit{MBR}}^{\mathcal{M}} = \sum_{i=0}^{\tau-2} \left( \Delta[(Q_i + PS) G R^i] + \nabla[(1-P)T_i G R^i]\right) + \nabla[L(n,\tau)]\\
        = \sum_{i=0}^{\tau-2} \left(\left \lceil \frac{4n}{gr^i c} \right \rceil \left \lceil \frac{GR^i}{q} \right \rceil \mathcal{A} P^i + \left \lceil \frac{GR^i}{q} \right \rceil S P^{i+1} + \left \lceil \frac{n^2 P^i (1-P)}{qc} \right \rceil \right)+ \mathcal{A}\left \lceil \frac{n^2}{qc} \right \rceil P^{\tau-1}.
\end{dmath}
In theory the MBR can potentially produce higher performance than SBR for the first levels of the subdivision process, because regions are naturally large in size and few in quantity. This is an scenario where MBR can employ all of its $q \times c$ resources efficiently, while SBR cannot. 
However, the GPU execution of an MBR scheme can introduce extra performance overhead, diminishing the benefit gained on these first levels of the sub-division process because of the additional block scheduling per region. When reaching deeper levels of depth there would exist sufficient regions to match or surpass the $q$ value, making SBR as competitive as MBR. 

\subsubsection{Theoretical Speedup of SBR/MBR Schemes}
\begin{figure*}[ht!]

    \centering
    \includegraphics[scale=0.58]{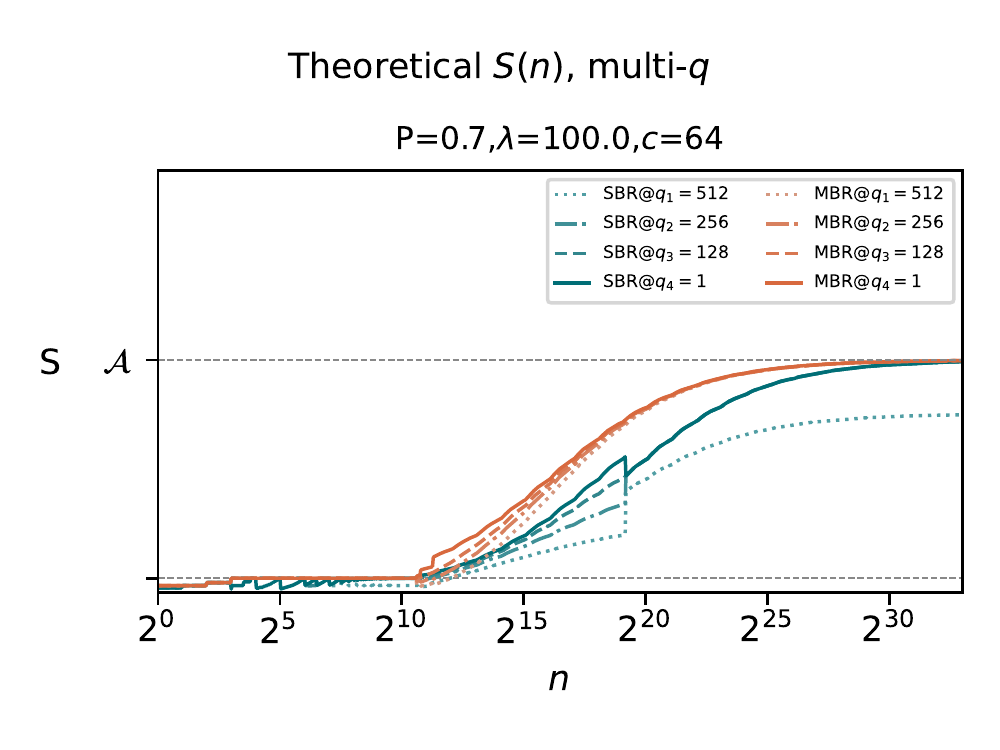}
    \includegraphics[scale=0.58]{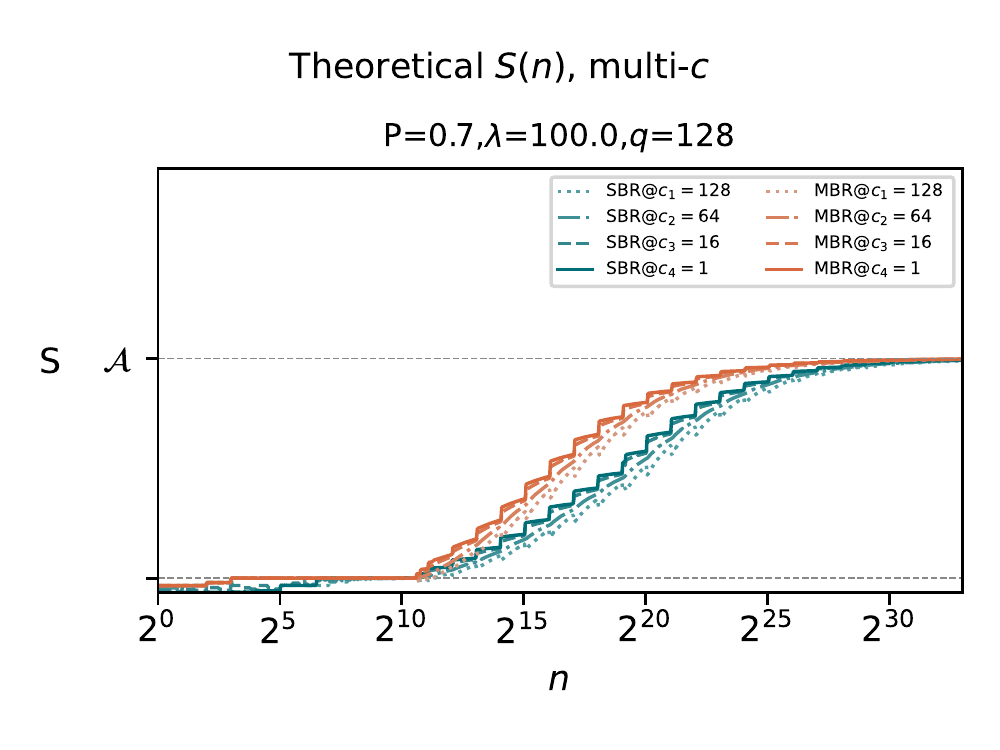}
    \includegraphics[scale=0.58]{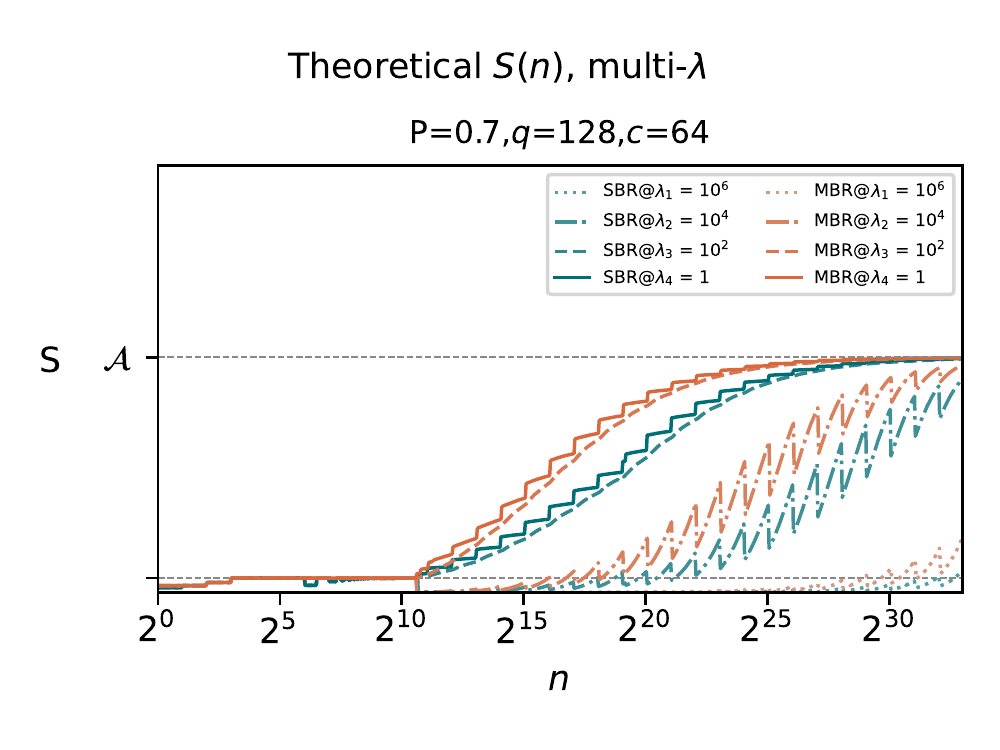}\\
    
    \includegraphics[scale=0.58]{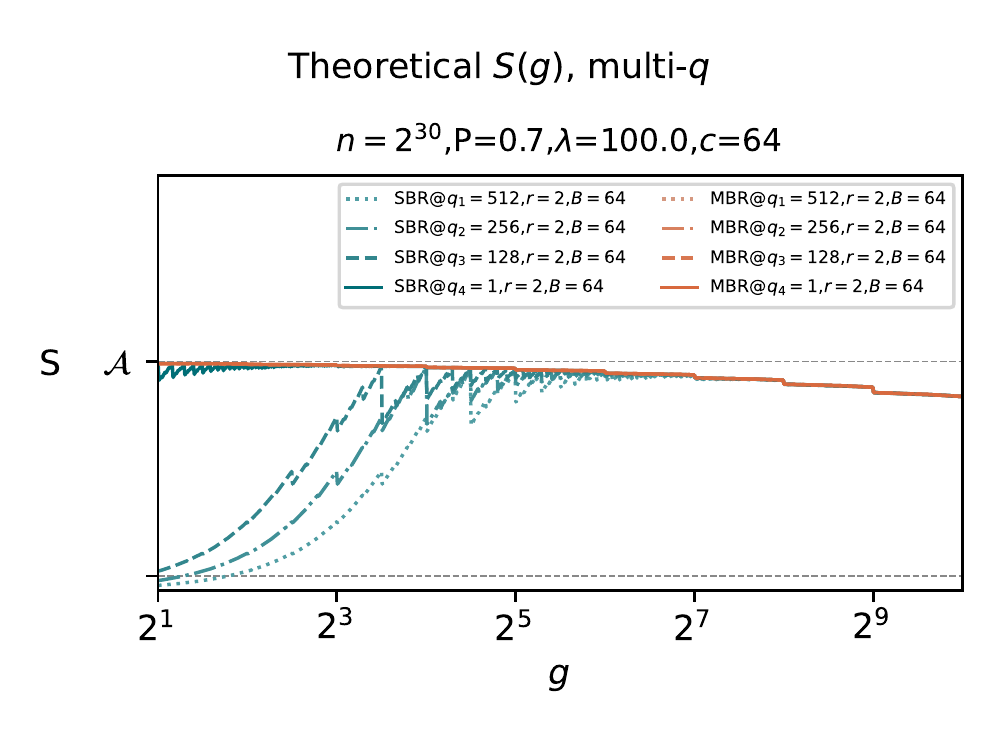}
    \includegraphics[scale=0.58]{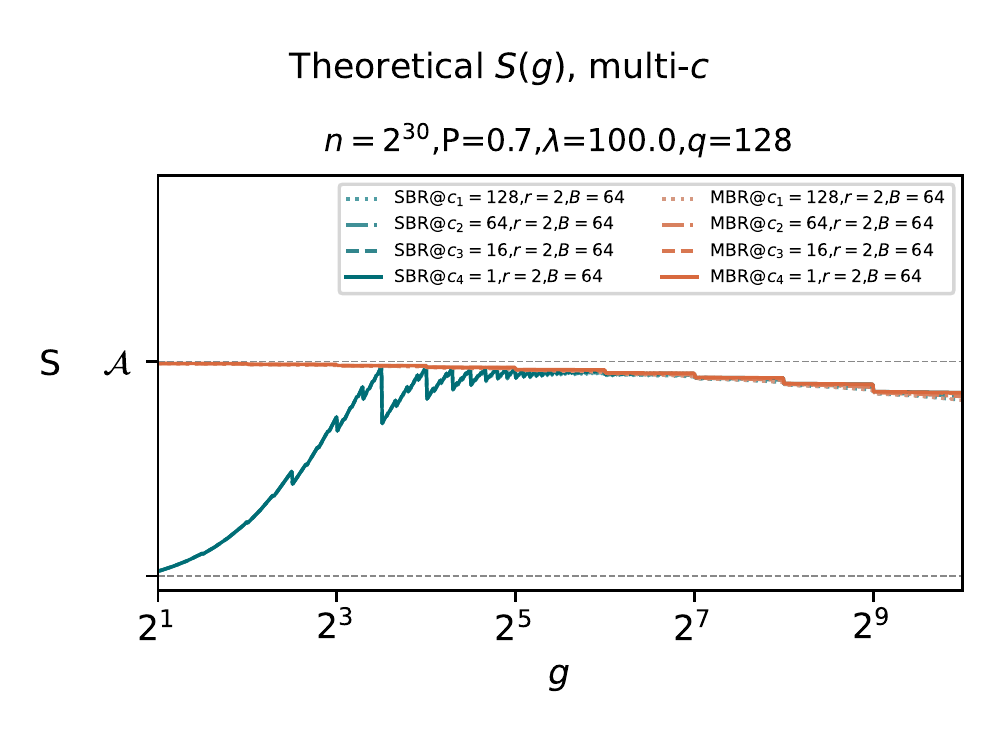}
    \includegraphics[scale=0.58]{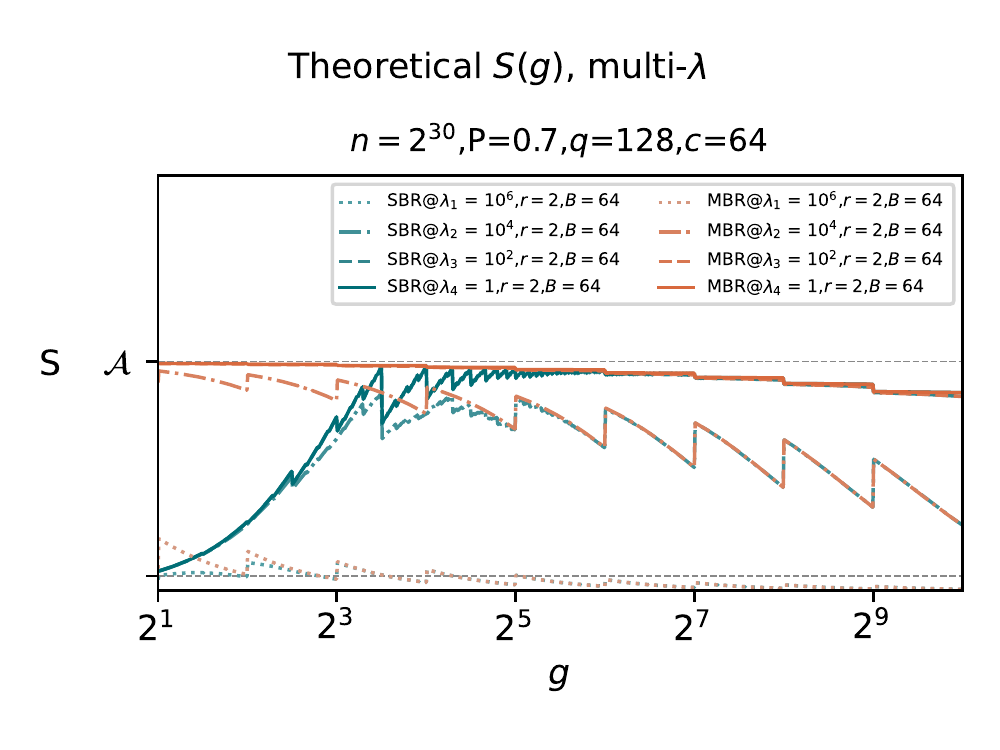}\\
    
    \includegraphics[scale=0.58]{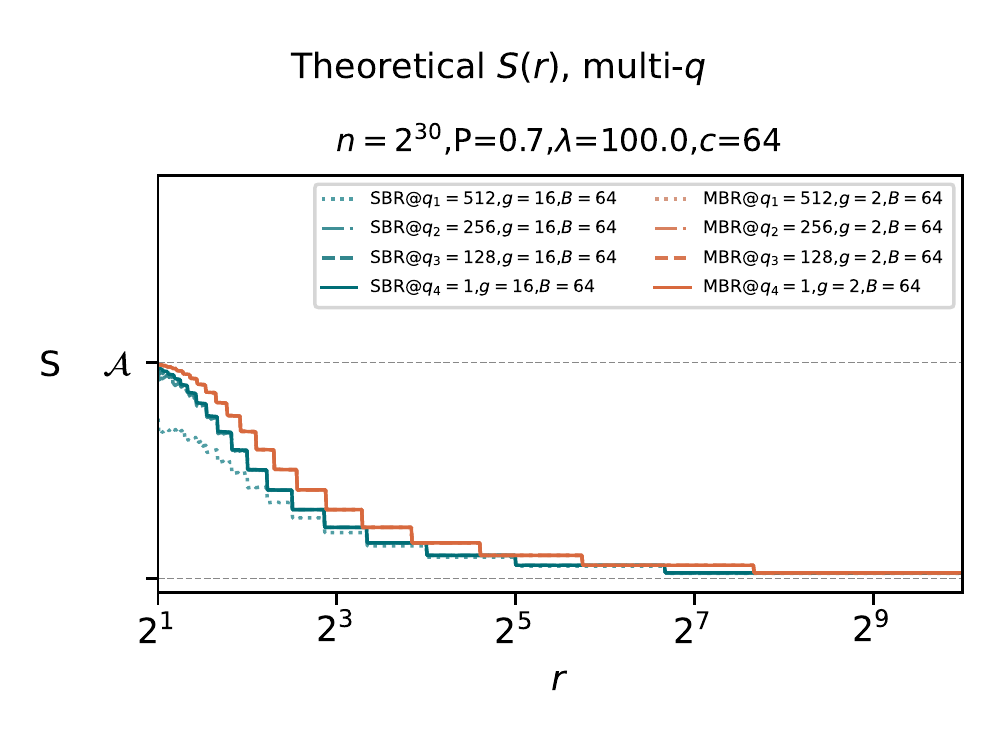}
    \includegraphics[scale=0.58]{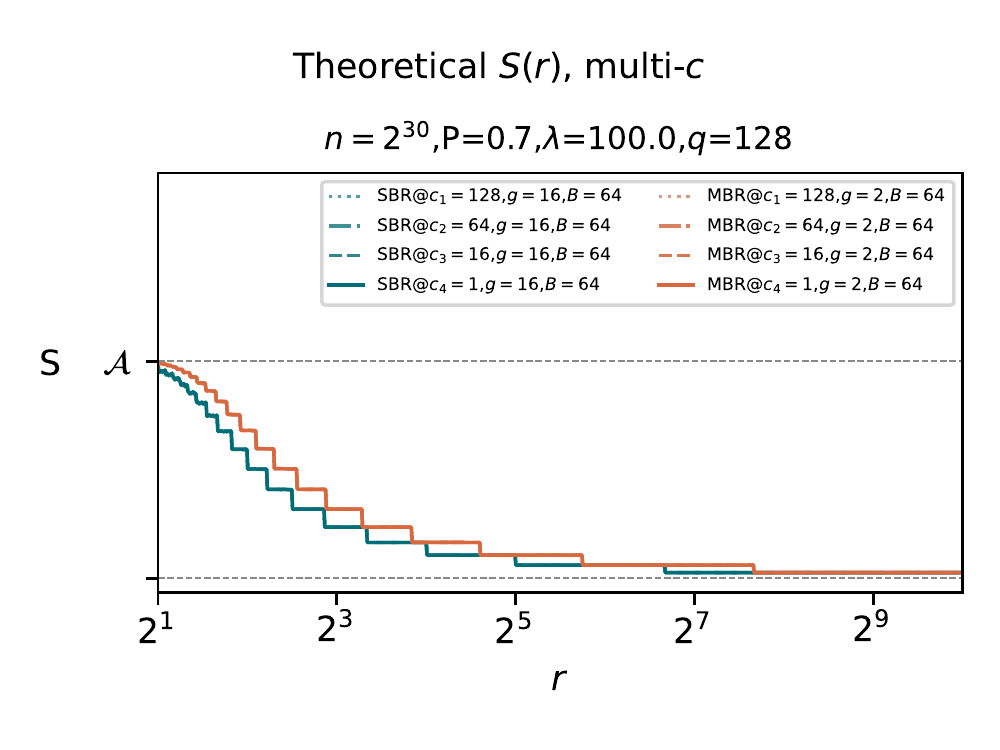}
    \includegraphics[scale=0.58]{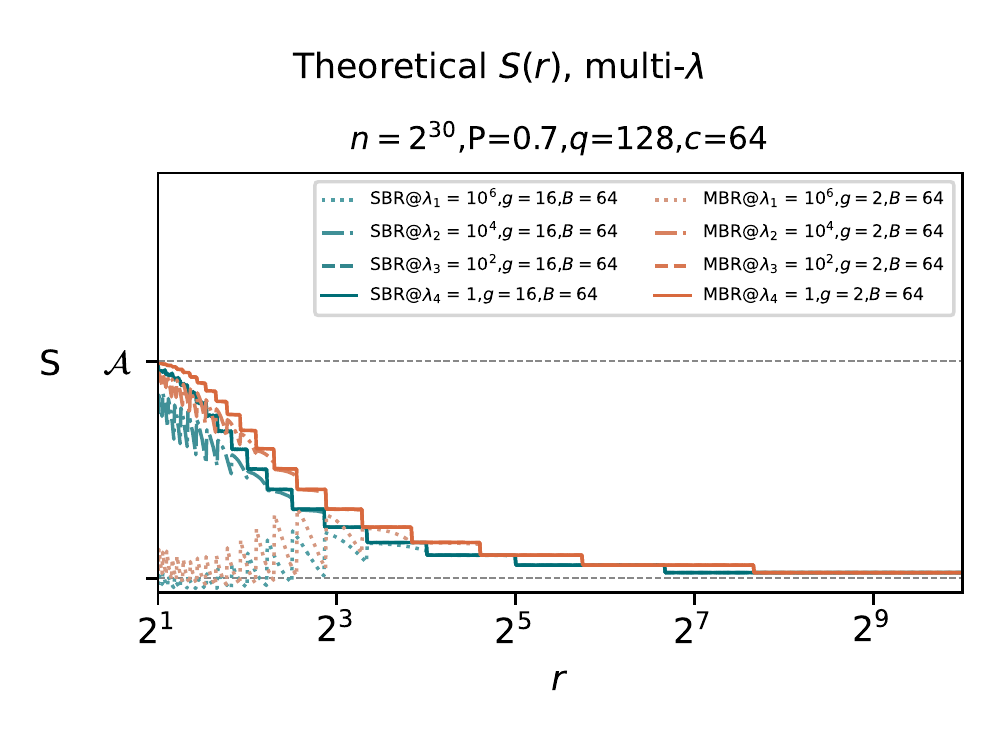}\\
    
    \includegraphics[scale=0.58]{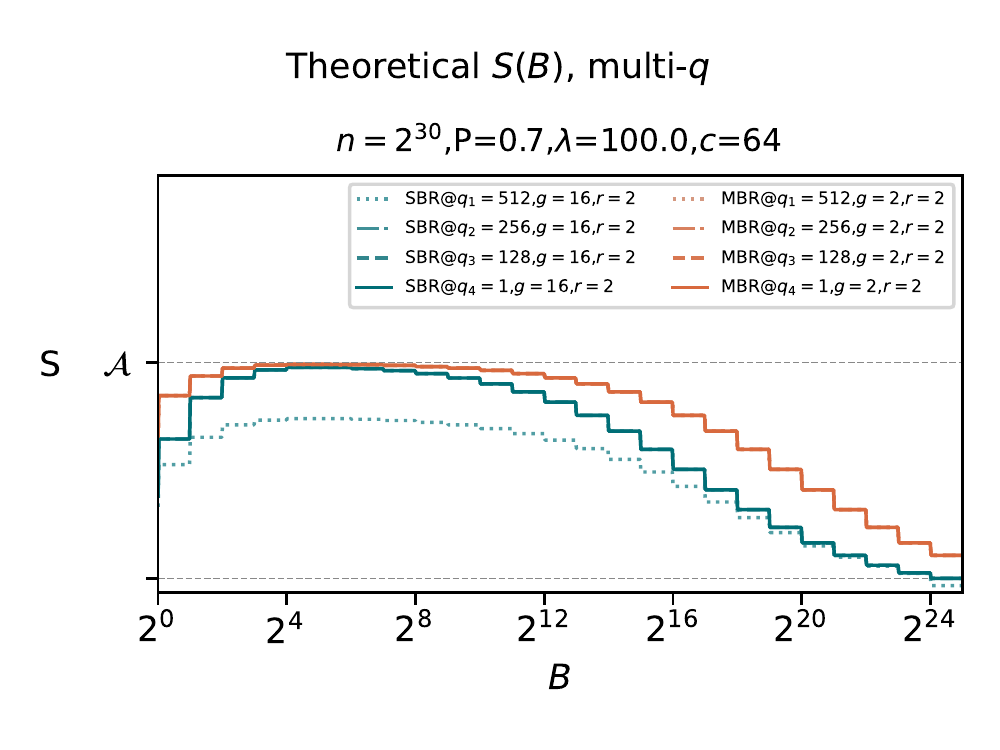}
    \includegraphics[scale=0.58]{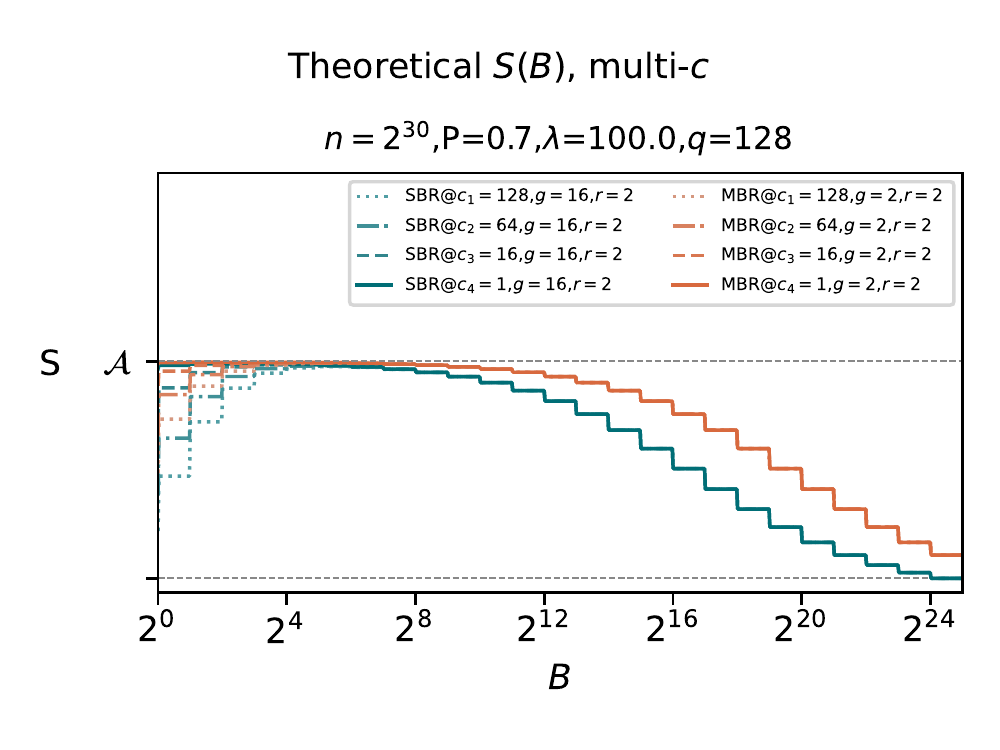}
    \includegraphics[scale=0.58]{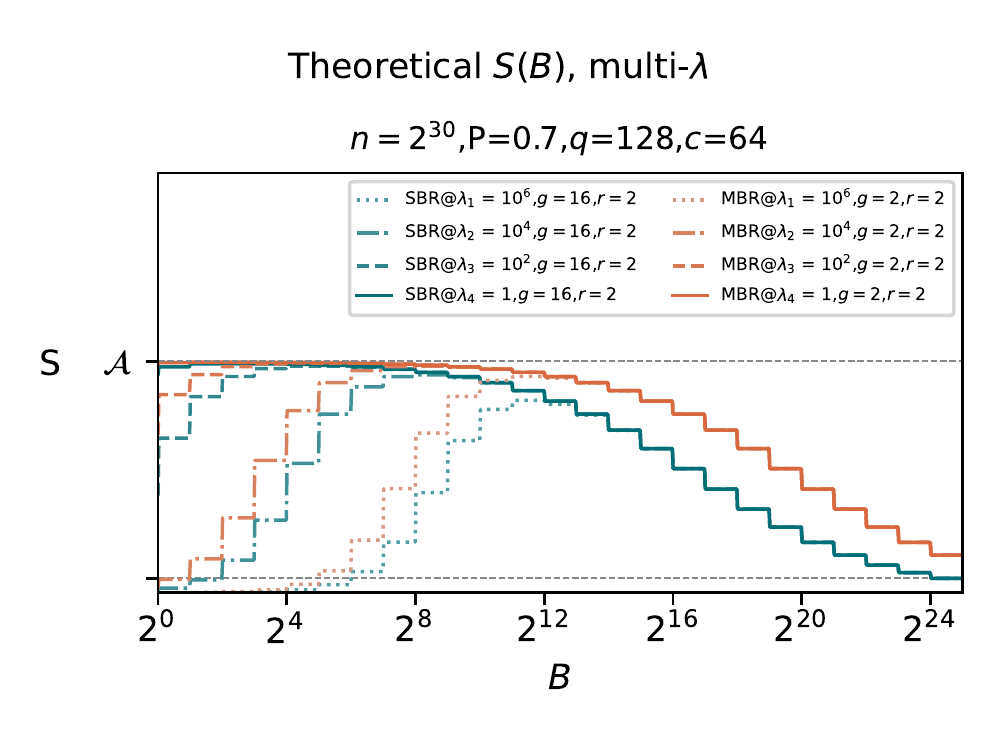}
    \caption{Theoretical Speedup using optimal $\{g,r,B\}$ parameters.}
    \label{fig:speedup-analytic}
\end{figure*}

The theoretical speedup is measured as the acceleration factor of the SBR/MBR schemes with respect to the Exhaustive (Ex) approach: 
\begin{align}
    \mathcal{S_{\textit{SBR}}} = \frac{T_{Ex}}{T_{\textit{SBR}}},\ \ \ 
    \mathcal{S_{\textit{MBR}}} = \frac{T_{Ex}}{T_{\textit{MBR}}}
\end{align}
Figure \ref{fig:speedup-analytic} presents the set of theoretical speedup plots using $q=128, c=64$ which represents a modern GPU in terms of the number of streaming-multiprocessors (SM) and the number of cores per SM, respectively. The plots explore the same parameters as in $\Omega$, with the addition of $S(q), S(c)$ to observe the impact of parallel resources on the speedup of each approach. As with $\Omega$, the $\mathcal{A}$ becomes an upper bound for speedup. 

The first row of plots, $S(n)$, shows that speedup becomes beneficial starting from $n \ge 2^{10}$ when $\lambda \le 100$, otherwise requires a larger problem size to be beneficial. Varying $q$ has a more negative impact on the SBR scheme, while varying $c$ has the same impact on both SBR/MBR schemes. 

The second row, $S(g)$, shows that for the MBR scheme the preferred initial subdivision is the smallest one $g \sim 2$. In the case of SBR, the scenario is different, as it suggests an initial subdivision sear $g \sim 2^5$. Varying $q$ shifts the optimal initial subdivision to the right, while varying $c$ does not have any significant effect. Varying $\lambda$ affects the MBR negatively, making it match the speedup of the SBR one for certain ranges of $g$.

The third row, $S(r)$, shows that for both SBR/MBR schemes, the highest speedup is achieved with a small recurrent subdivision, such as $r \sim 2$. Variations of $q$ and $c$ do not have any significant impact on the optimal value of $r$. On the other hand, large values of $\lambda$ can shift the optimal $r$ value to the right, such as $\lambda \sim 10^6$ that shifts the optimal recurrent subdivision to $r \sim 2^3$.

The fourth row, $S(B)$, shows that the optimal values for $B$ have slightly shifted to the right in comparison to $\Omega(B)$, now suggesting an optimal of $B \sim 2^5$. Varying $q$ impacts negatively on SBR, while varying $c$ does not show a significant impact. Varying $\lambda$ shows that higher subdivision costs shift the optimal $B$ to larger values which is logical in order to reduce the total number of subdivisions. 

The theoretical plots of $S(n), S(g), S(r), S(B)$ have shown relevant behaviors for SBR and MBR under the different $\{g,r,B\}$ parameters. Typically, the sub-division approach would be handled by Dynamic Parallelism, but it could be handled by an alternative iterative approach as well. In that regard, it would be useful to know how each the SBR and MBR schemes perform both in a recursive and iterative approach. In the next Section we present \textit{Adaptive Serial Kernels} as the iterative version of Dynamic Parallelism, which is also used later for experimental comparisons with DP. 

\section{Adaptive Serial Kernels (ASK)}
\label{sec:ask-formulation}
\subsection{General View of ASK}
\begin{figure}[ht!]
    \centering
    \includegraphics[scale=0.6]{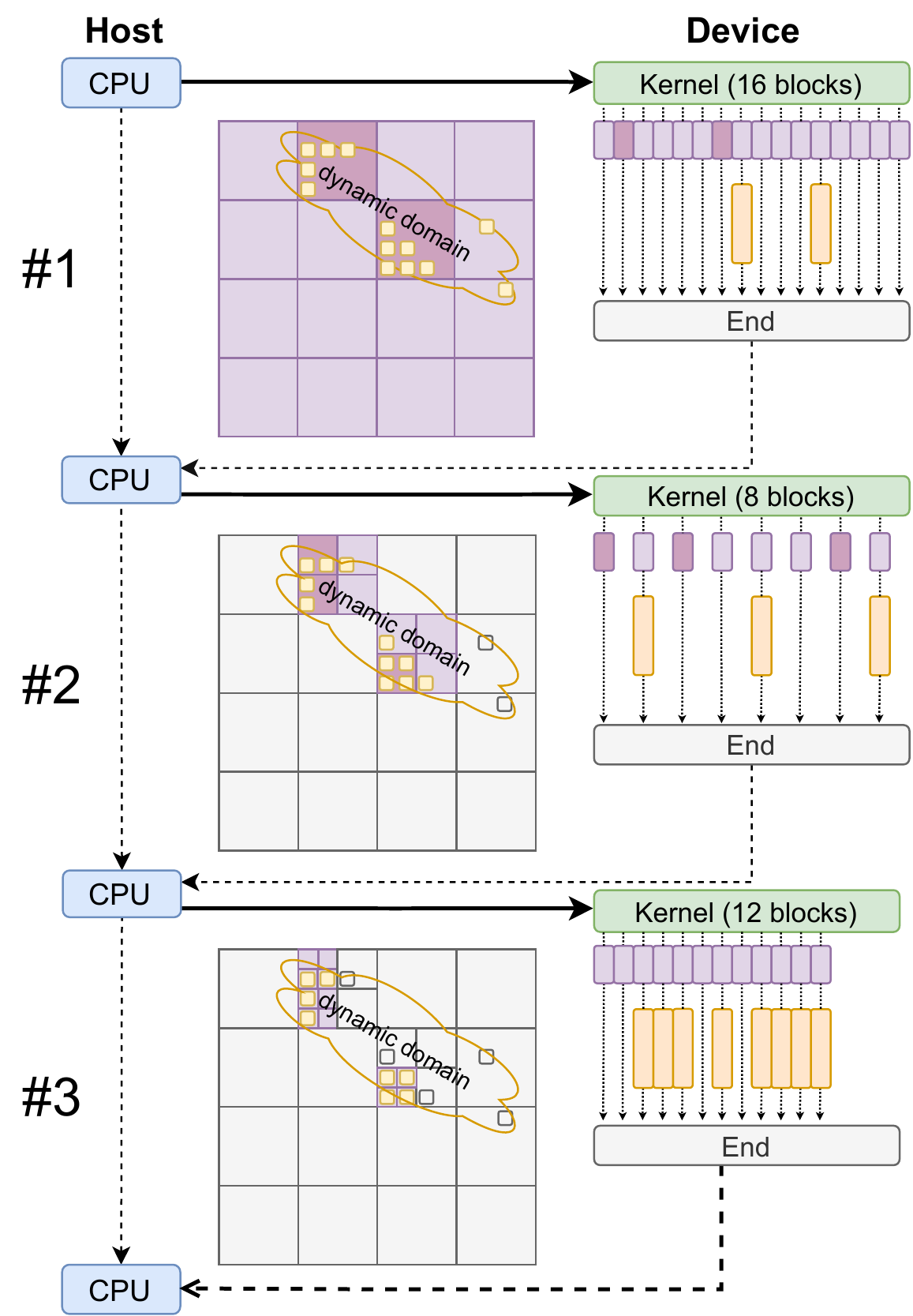}
    \caption{Adaptive Serial Kernels (ASK) for an example with three iterations. Light purple corresponds to exploration with no subdivision. Dark purple corresponds to exploration that did found further subdivision. The vertical yellow bars correspond to application work ($\mathcal{A}$). The initial grid has $|G_x| \times |G_y| = 4 \times 4$ regions with subdivisions of $r_x \times r_y = 2 \times 2$ at each iteration. }
    \label{fig:iterative-approach-general-view}
\end{figure}

The intuition behind ASK is to replace the recursive solution design used in DP by an iterative approach where a small number of flat kernels are executed serially one after another, with just the necessary number of parallel resources at each iteration. The relevance of ASK in the context of the subdivision cost model presented in Section \ref{sec:work-model} is that there is an opportunity for ASK to introduce a lesser subdivision overhead $S$ and improve CUDA's DP performance, because kernels would be called one per depth level, instead of one kernel per node of the subdivision tree as in DP.  
Figure \ref{fig:iterative-approach-general-view} illustrates the process of ASK in the case of a SBR scheme. 

In the example an initial compute grid of $|G_0| = 4 \times 4$ thread-blocks is applied. On the rest of the kernel calls, the grid is dynamically defined in terms of the processing and subdivision process of the previous kernel call. Each thread-block may subdivide into $r_x \times r_y = 2 \times 2$ blocks if further parallelism is required at that region, otherwise a final terminal pass is done, if required. Given that it is a SBR scheme, regions are handled by a single GPU thread-block allowing thread collaboration and synchronization within each region. 
\subsection{Offset Lookup Tables (OLTs)}
In order to keep track of the active regions at every iteration, ASK keeps two offset-lookup-tables (OLT) updated; a read-OLT for accessing the region coordinates of the current level, and a write-OLT for writing the new sub-region coordinates of every region, as the result of subdivision. Whenever a thread-block subdivides, one of its threads writes down into the write-OLT, from left to right and concurrently with the other thread-blocks, the $r_x \times r_y$ coordinate offsets of the new generated thread-blocks that will be used on the next iteration. OLTs are accessed by their block identifier, thus it allows a pure data-parallel access pattern for the blocks of the upcoming grid in the next iteration. In terms of memory, the size of read and write OLTs is compact and corresponds to one $(x,y)$ coordinate per active regions, \textit{i.e.}, it does not need to store past finalized regions neither allocate extra storage for future regions. 
Lastly, the OLT can be kept in device memory throughout all of the iterations without needing to be copied back to host in-between kernels. 
One key aspect of ASK is to perform parallel operations on the OLT efficiently. The next sub-section explains the process in detail.

\subsection{Operations on the OLT} 
Handling the offsets lookup table (OLT) introduces two main challenges: i) to achieve compact concurrent insertions and ii) to keep the OLT's size close to a minimum. 

\subsubsection{Compact Concurrent Insertions on the OLT} 
Insertions in the OLT occur when a region subdivides into $r_x \times r_y$ sub-regions concurrently with the other thread-blocks that are handling other active regions. These new sub-region positions are inserted in the write-OLT satisfying two requirements: i) to occur concurrently for all thread-blocks that subdivide in the current iteration and ii) all insertions should end up in a compact form, \textit{i.e.}, no empty spaces in the write-OLT. These two requirements can be satisfied by using an atomic-add operation, which offers consistent concurrent addition on a variable shared by multiple threads of different thread-blocks, and also returns the previous value it had for each thread that uses the operation. Therefore, when a block decides to subdivide a region, its thread in charge atomically increments a \textit{count} GPU-global variable by one and obtains its previous value as well. The previous value of \textit{count} corresponds to the position in the write-OLT where there are $r_x \times r_y$ consecutive slots to write the new sub-block offsets. By being an atomic operation, the concurrent operations performed on the variable produce a compact insertion scheme which is also free of race conditions.
Another benefit of this process is that once the current kernel finalizes (thus all the corresponding subdivisions have been done for that iteration) the \textit{count} variable will contain the total number of regions to subdivide, which corresponds to the number of blocks (size of the grid) to use for the next kernel iteration. 

The atomic-add process could also be replaced by combining per-block prefix-sum ($\mathcal{O}(\log(n))$ time) with global atomic-add in case the pure atomic-add approach is not as fast as expected. A full prefix-sum is less preferred because the synchronization between blocks requires special treatment, such as collaborative groups at the cost of a limit in problem size, or multiple kernel passes at the cost of extra overhead.

\subsubsection{Swapping and Keeping the OLT Size to a Minimum}
At each iteration, the read-OLT and write-OLT swap their roles. That is, at the next iteration, the write-OLT becomes the read-OLT, and the read-OLT is now a write-OLT with the necessary memory reallocated to store the potential new subdivisions of all current active regions (\textit{i.e.}, the regions specified in the current read-OLT). The memory reallocation of the new write-OLT uses the \textit{count} value to know how many regions will be processed in the next execution. The size of the write-OLT is $count \times (r_x \cdot r_y)$, where $count$ is the number of active regions at a given iteration, and $(r_x \cdot r_y)$ is the extra allocated space for the potential subdivisions at every active region.

\section{Experimental Case Study: The Mandelbrot Set}
\label{sec:experimental-results}
Experimental tests are performed using the \textit{Mandelbrot Set} as the case study, which defines a set of complex numbers \textit{c} such that when passed into the dynamical system $z_{i+1} = z_i^2  + c$, with $z_0=0$, they satisfy $|z_{i \mapsto \infty}| \le 2$. The term \textit{dwell} refers to the number of iterations on the dynamical system before checking if $|z_{dwell}| < 2$ satisfies or not. An exhaustive algorithm performs this process on every pixel of the discrete complex plane.

\begin{figure}[ht!]
    \centering
    \includegraphics[scale=0.025]{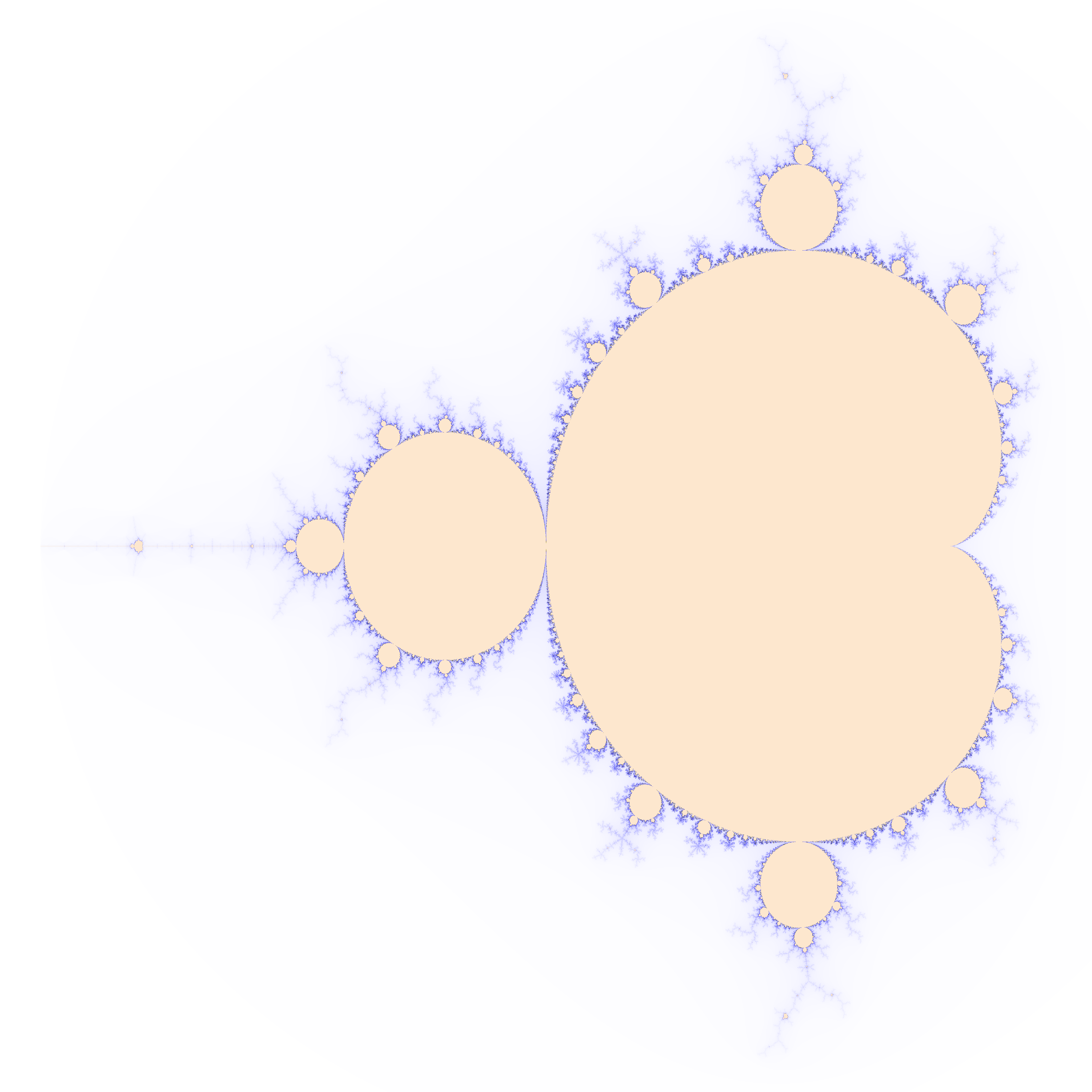}\ \ \ \ \ \ \ \ \ \ \ \ \ \
    \includegraphics[scale=0.025]{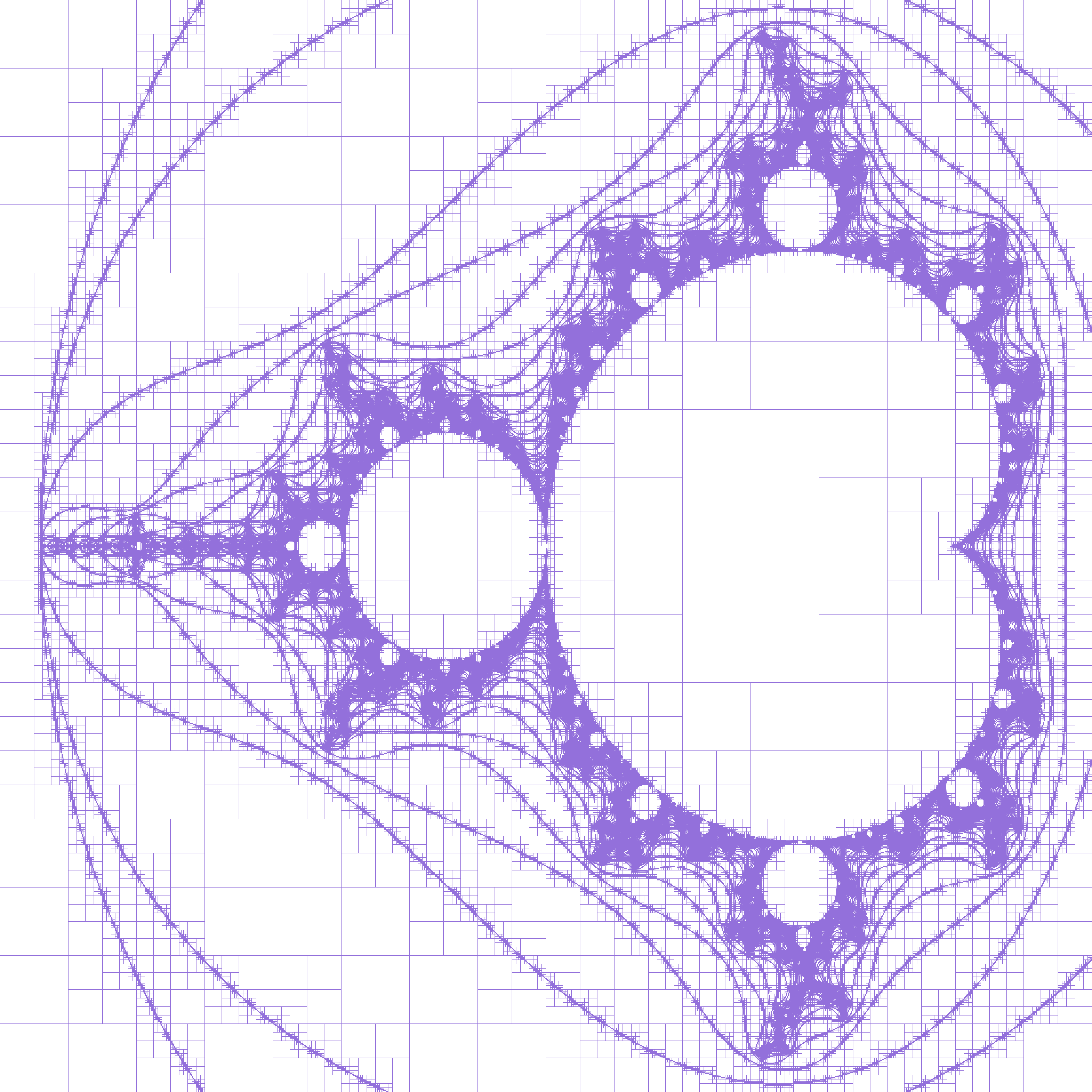}
    \caption{On the left, the Mandelbrot set. On the right, the dynamic grid of regions generated through subdivision, revealing the fractal's geometry.}
    \label{fig:mandelbrot}
\end{figure}

The Mariani-Silver algorithm \cite{mariani-silver} is a more efficient approach, that through subdivision, can detect if a given region can be discarded entirely or not, avoiding unnecessary work to be done. Fig. \ref{fig:mandelbrot} (left) shows an illustration of the Mandelbrot Set and its resulting grid (right) after several subdivision with the Mariani-Silver algorithm.

\subsection{Implementations and Experimental Design}
\label{subsec:experimental-design}
Three implementations are put to test in terms of performance:
\begin{enumerate}
    \item \textbf{Ex}: Exhaustive approach in one flat kernel execution.
    \item \textbf{DP}: Recursive kernels with CUDA Dynamic Parallelism.
    \item \textbf{ASK}: Proposed iterative alternative to CUDA DP.
\end{enumerate}
 The DP and ASK approaches subdivide into two variants of the Mariani-Silver algorithm; single-block per Region (SBR) and multiple-blocks per region (MBR) schemes. The DP implementation was developed by Adinetz \cite{adinetz-dp} from Nvidia, that uses the DP-MBR scheme. The DP-SBR variant is a modification of Adinetz's implementation at the subdivision stage in order to employ one block per region. The ASK-SBR/MBR variants were implemented by the authors. The source code of these approaches, and its benchmark, are available in Github\footnote{\url{https://github.com/temporal-hpc/GPU-dynamic-parallelism}}.

The performance test consists of measuring the average execution time on generating the Mandelbrot set in the complex plane $[-1.5, -1]\times[0.5, 1]$, using a dwell of $d=512$. The problem size $n \times n$ varies in the range $n \in [1024..65536]$ or up to the largest $n$ that fits in GPU memory. The average execution time was computed from $100$ realizations, where each program execution returns a sub-average value from $20$ internal repetitions. With these settings, the standard error remained less than $1\%$. The performance tests were conducted on four different GPUs which are detailed in Table \ref{tab:hardware}. Programming and testing was done on Linux with CUDA C/C++ API \cite{cudaCProgGuide2021}. 

\begin{table*}[ht!]
\caption{Hardware specifications of the GPUs and computer systems used for the experiments.}
\label{tab:hardware}
\resizebox{\textwidth}{!}{
\begin{tabular}{|ll|ll|l|l|l|}
\hline
\multicolumn{2}{|l|}{\textbf{\cellcolor{temporal}Attribute}}                   & \multicolumn{2}{l|}{\cellcolor{temporal}\textbf{Nvidia A100}}      & \cellcolor{temporal}\textbf{Nvidia TITAN RTX} & \cellcolor{temporal}\textbf{Nvidia TITAN V} & \cellcolor{temporal}\textbf{Nvidia Jetson NX}    \\ \hline
\multicolumn{2}{|l|}{\textbf{Segment}}                                        & \multicolumn{2}{l|}{HPC Server (DGX A100)}     & Workstation               & Workstation             & Embedded                     \\ \hline
\multicolumn{2}{|l|}{\textbf{Architecture}}                                   & \multicolumn{2}{l|}{Ampere}                    & Turing                    & Volta                   & Volta                        \\ \hline
\multicolumn{2}{|l|}{\textbf{GPU Chip}}                                       & \multicolumn{2}{l|}{GA100}                     & TU102                     & GV100                   & GV10B                        \\ \hline
\multicolumn{2}{|l|}{\textbf{Cores (FP32)}}                                     & \multicolumn{2}{l|}{$6912$}                    & $4608$                    & $5120$                  & $384$                        \\ \hline
\multicolumn{2}{|l|}{\textbf{SMs ($q$ value)}}                                     & \multicolumn{2}{l|}{$108$}                    & $72$                    & $80$                  & $6$                        \\ \hline
\multicolumn{2}{|l|}{\textbf{Cores/SM ($c$ value)}}                                     & \multicolumn{2}{l|}{$64$}                    & $64$                    & $64$                  & $64$                        \\ \hline
\multicolumn{2}{|l|}{\textbf{Memory}}                                         & \multicolumn{2}{l|}{$40GB$}                    & $24GB$                    & $12GB$                  & $8GB$ (shared with CPU)               \\ \hline
\multicolumn{2}{|l|}{\textbf{Memory Bandwidth}}                                      & \multicolumn{2}{l|}{$1.5GB/s$}                & $672GB/s$                & $651.3GB/s$            & $51.2GB/s$                  \\ \hline
\multicolumn{2}{|l|}{\textbf{Max Power (W)}}                                  & \multicolumn{2}{l|}{$400W$}                    & $280W$                    & $250W$                  & $20W$                        \\ \hline
\multicolumn{1}{|l|}{\multirow{3}{*}[-0.5em]{\textbf{System}}} & \textbf{OS}  & \multicolumn{2}{l|}{DGX OS 5.2 (Ubuntu Based)}                & Arch Linux                & Arch Linux              & Ubuntu 18.04                       \\ \cline{2-7} 
\multicolumn{1}{|l|}{} & \textbf{CUDA}  & \multicolumn{2}{l|}{11.4.2}                &  11.6                & 11.6              & 10.2                       \\ \cline{2-7} 
\multicolumn{1}{|l|}{}                                         & \textbf{CPU} & \multicolumn{2}{l|}{2 x 64-core AMD EPYC 7742} & Intel 10-core i7-6950X    & Intel 10-core i7-6950X  & Nvidia 6-core Carmel ARMv8.2 \\ \cline{2-7} 
\multicolumn{1}{|l|}{}                                         & \textbf{RAM} & \multicolumn{2}{l|}{1 TB RAM DDR4-3200Hz}      & 128GB DDR4 2400Mhz        & 128GB DDR4 2400Mhz      & 8GB LPDDR4x 1600Mhz          \\ \hline
\end{tabular}
}
\end{table*}

\subsection{Finding Optimal blocksizes and $\{g,r,B\}$ configurations}
Different blocksizes of $8\times 8$, $16\times 16$, $32\times 32$, $64\times 4$ and $64\times 8$ were explored in order to find the best CUDA blocksize for each approach, on each tested GPU. The best block was chosen in terms of their fastest running times achieved for the largest problem sizes. Table \ref{tab:best-blocksizes} presents the chosen configurations.
\begin{table}[ht!]
\centering
\caption{Best CUDA blocksizes}
\label{tab:best-blocksizes}
\resizebox{\columnwidth}{!}{
\begin{tabular}{|r|r|r|r|r|}
\hline
        & A100  & TITAN RTX & TITAN V & JETSON XAVIER NX \\ \hline
Ex      & 16x16 & 8x8       & 16x16   & 64x4             \\ \hline
DP-SBR  & 64x8  & 64x8      & 64x8    & 16x16            \\ \hline
DP-MBR  & 64x8  & 64x8      & 64x8    & 16x16            \\ \hline
ASK-SBR & 16x16 & 16x16     & 16x16   & 8x8              \\ \hline
ASK-MBR & 16x16 & 64x4      & 64x4    & 16x16            \\ \hline
\end{tabular}
}
\end{table}

\begin{figure*}
    \label{fig:best-grB-configs}
    \centering
    \includegraphics[scale=0.445]{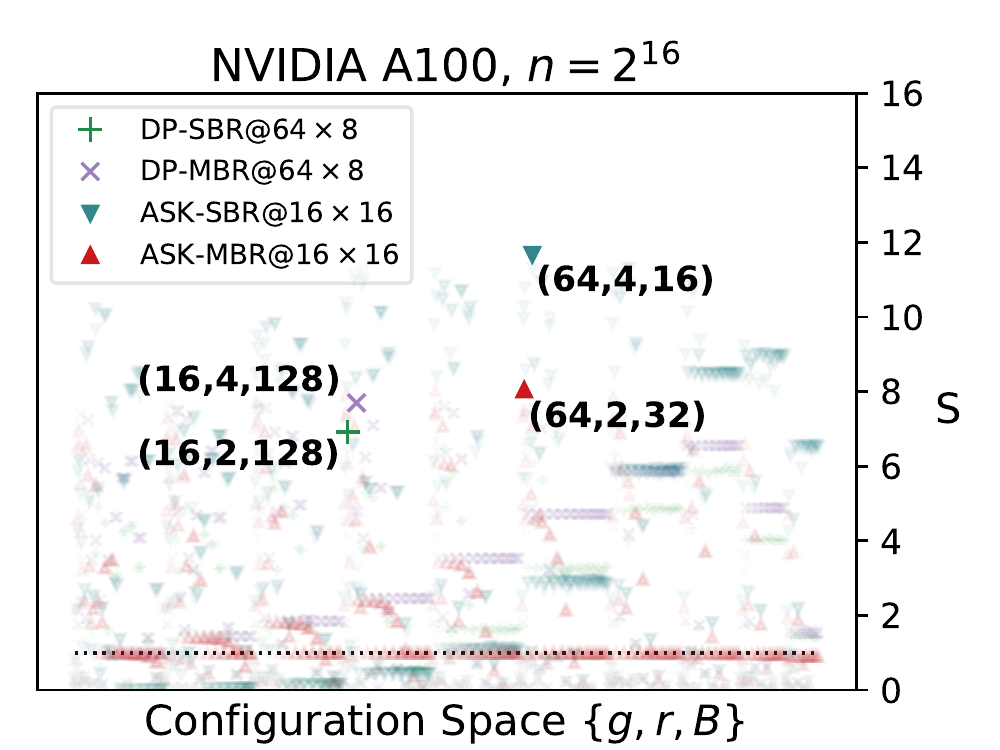}
    \includegraphics[scale=0.445]{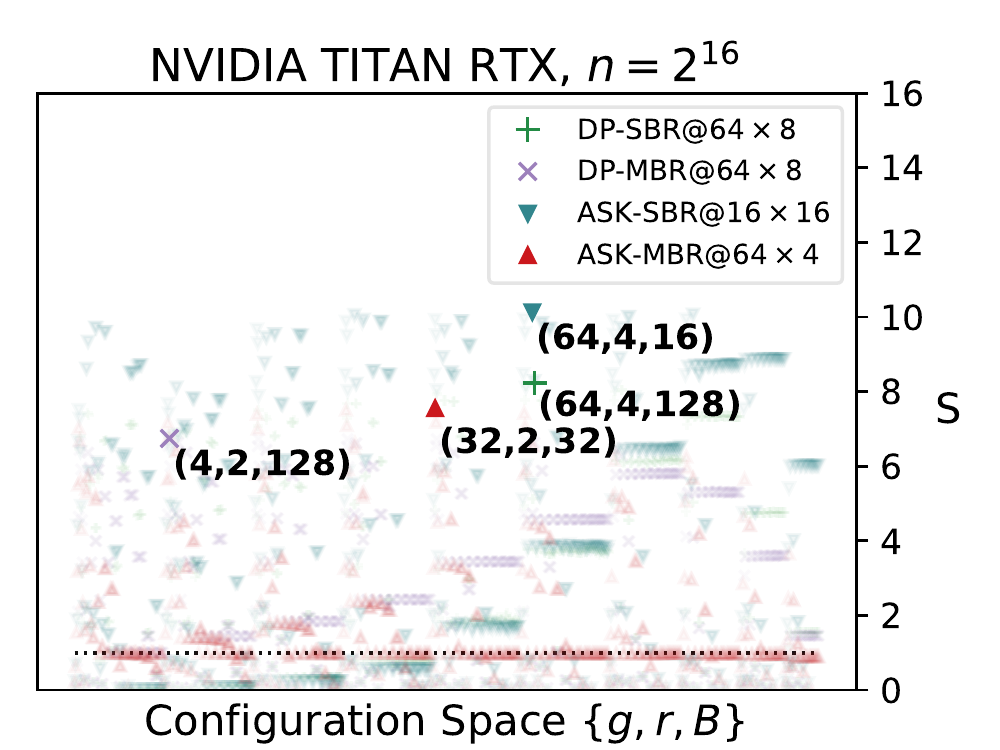}
    \includegraphics[scale=0.445]{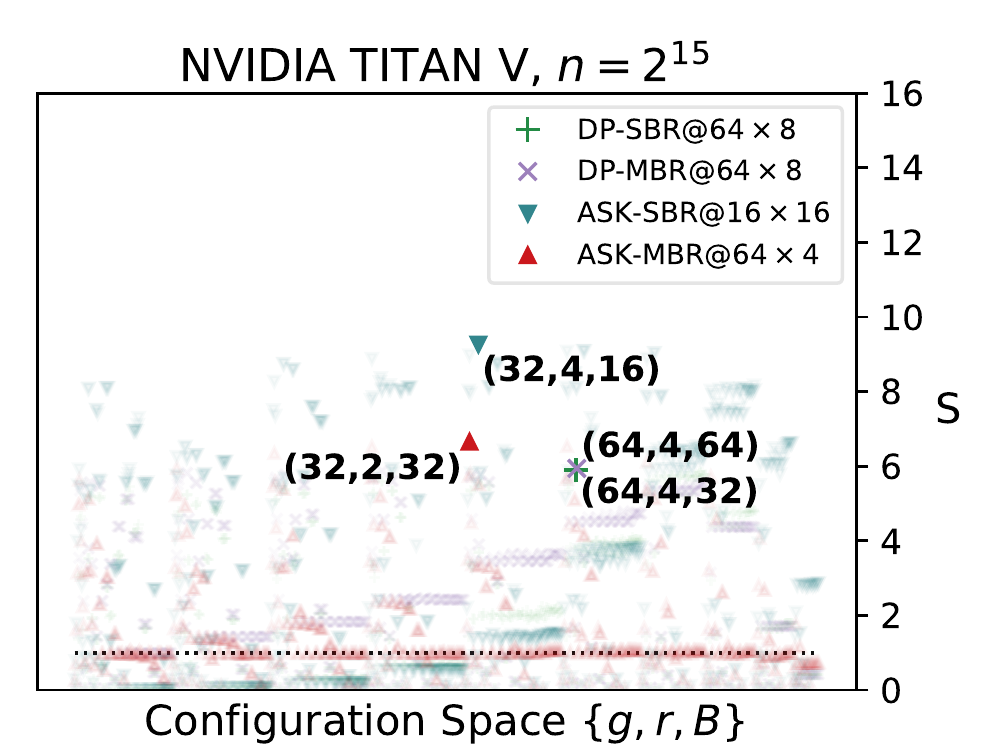}
    \includegraphics[scale=0.445]{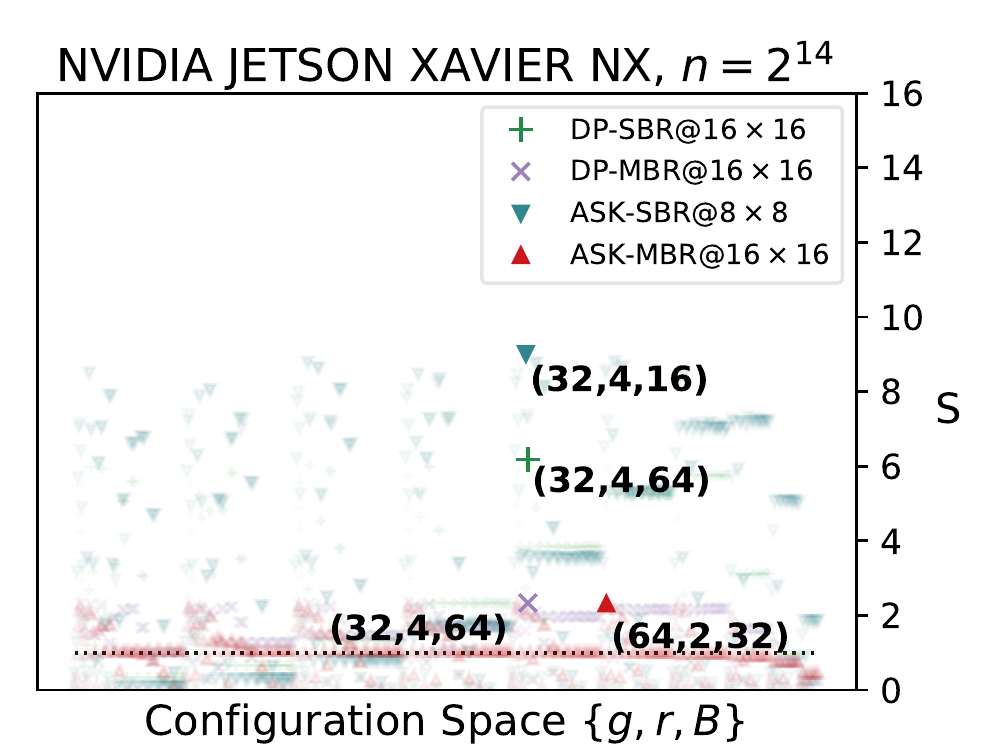}
    \caption{Speedup landscape of the DP and ASK approaches over Ex, using the configuration space $\{g,r,B\}$.}
\end{figure*}

The $\{g,r,B\}$ configuration space was also explored in the range $g,r,B \in \{2,4,8,\dots,1024\}$ for all approaches using the largest problem sizes each GPU allowed. Figure \ref{fig:best-grB-configs} presents the configuration landscapes in terms of speedup over the Exhaustive approach, using their corresponding optimal blocksizes and highlighting the optimal $\{g,r,B\}$ configuration for each approach.
From the landscapes, one can note that the optimal $\{g,r,B\}$ configuration changes for each approach, and varies on each different GPU. In the case of DP-SBR, its $g$ parameter may take the values $g=16,32,64$ depending of which GPU is used, its $r$ parameter remains stable and its $B$ parameter varies between $B=64,128$. The scenario is similar for the DP-MBR approach, with the exception of its $g=4$ value when ran on the TITAN RTX GPU. In the case of the ASK approach, the schemes present stable optimal configurations across the GPUs; the ASK-SBR scheme has an optimal configuration of $\{g,r,B\}=\{32/64, 4, 16\}$ while the ASK-MBR scheme has an optimal of $\{g,r,B\} = \{32/64, 2, 32\}$.  The performance results of the following sub-section use these optimal CUDA blocksizes and $\{g,r,B\}$ configurations. 

\subsection{Performance Results}
\label{subsec:results}
\begin{figure*}[ht!]

    \centering
    \includegraphics[scale=0.44]{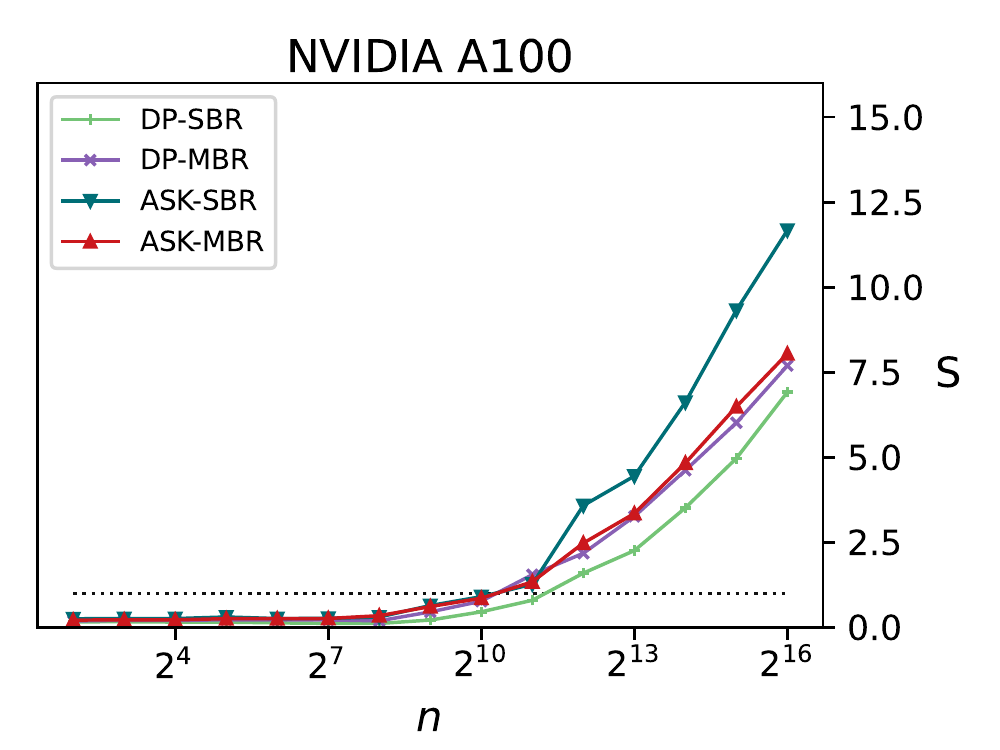}
    \includegraphics[scale=0.44]{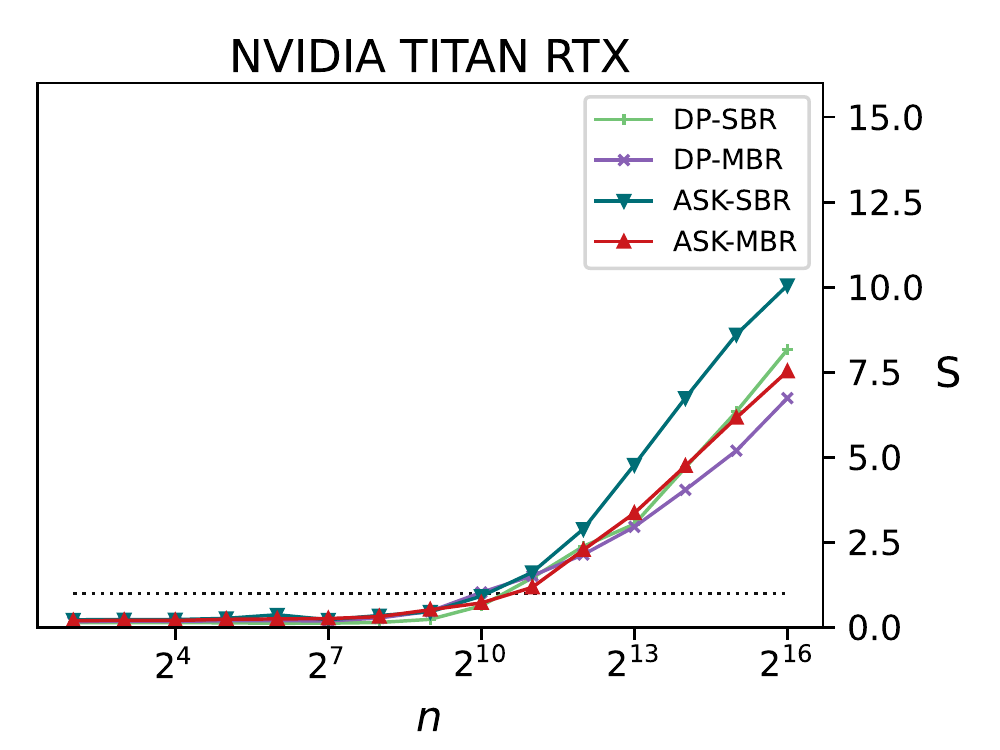}
    \includegraphics[scale=0.44]{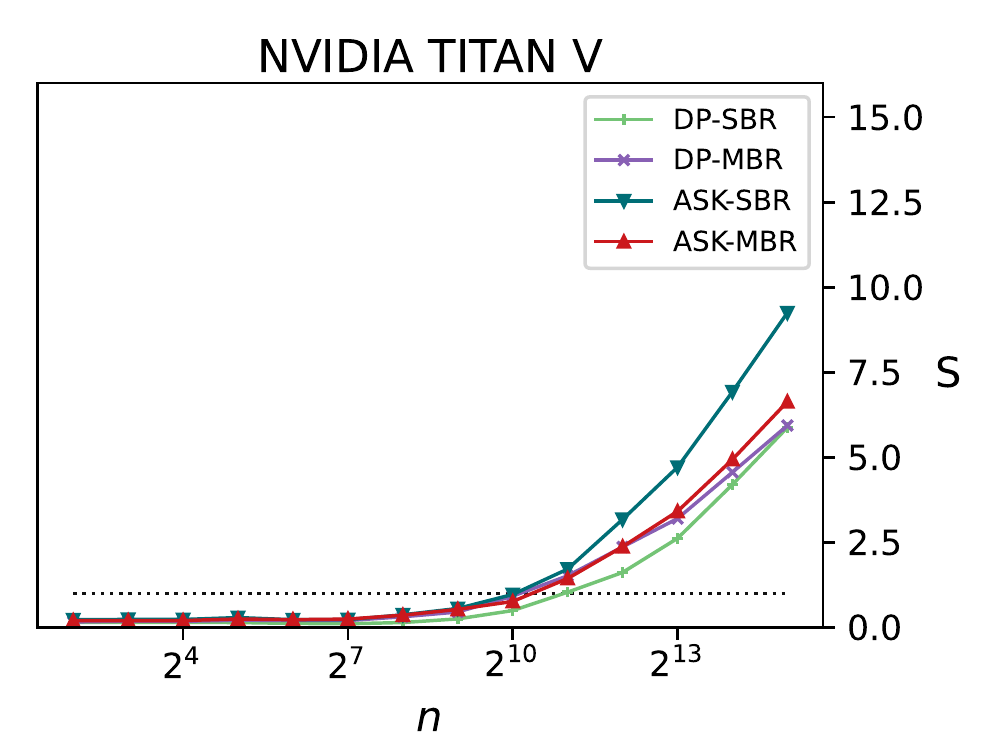}
    \includegraphics[scale=0.44]{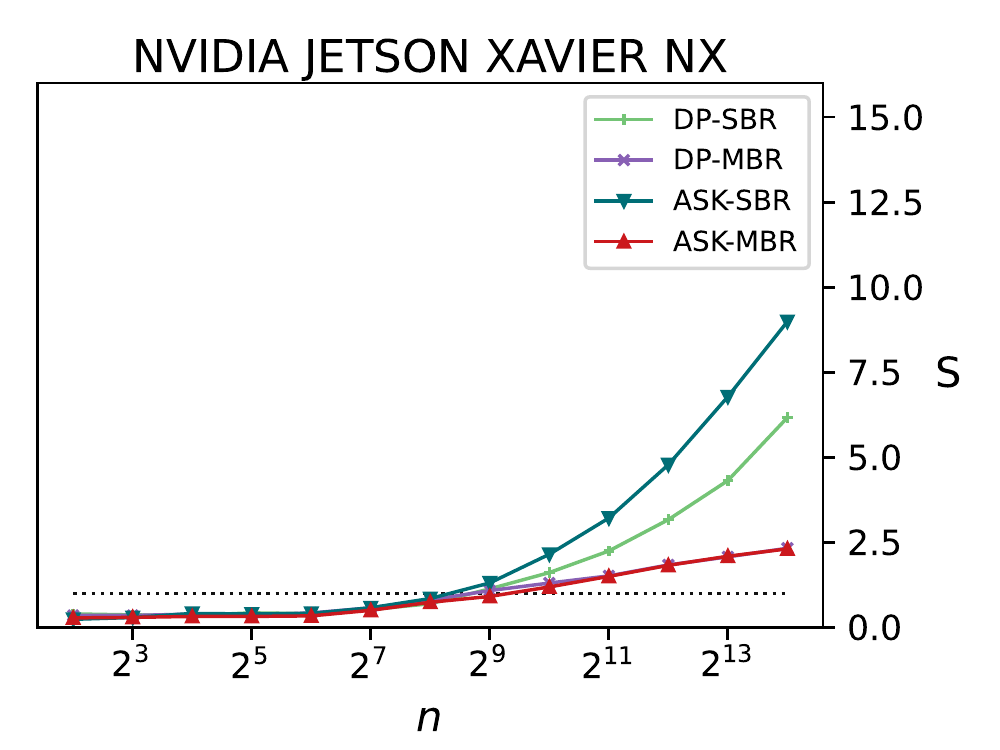}\\
    
    \includegraphics[scale=0.44]{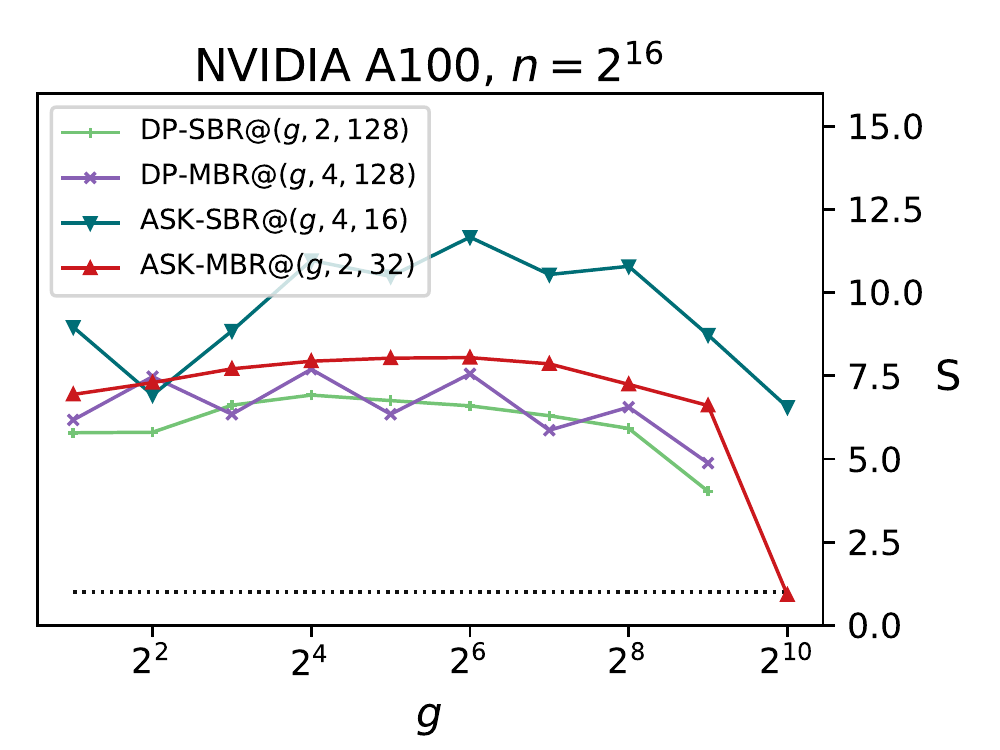}
    \includegraphics[scale=0.44]{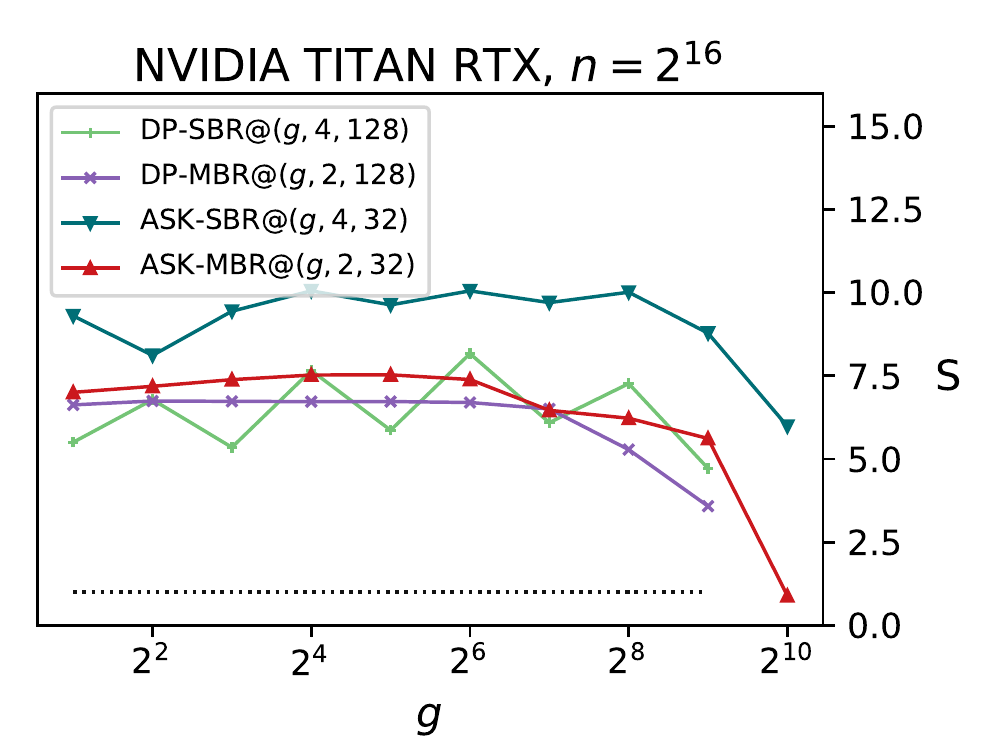}
    \includegraphics[scale=0.44]{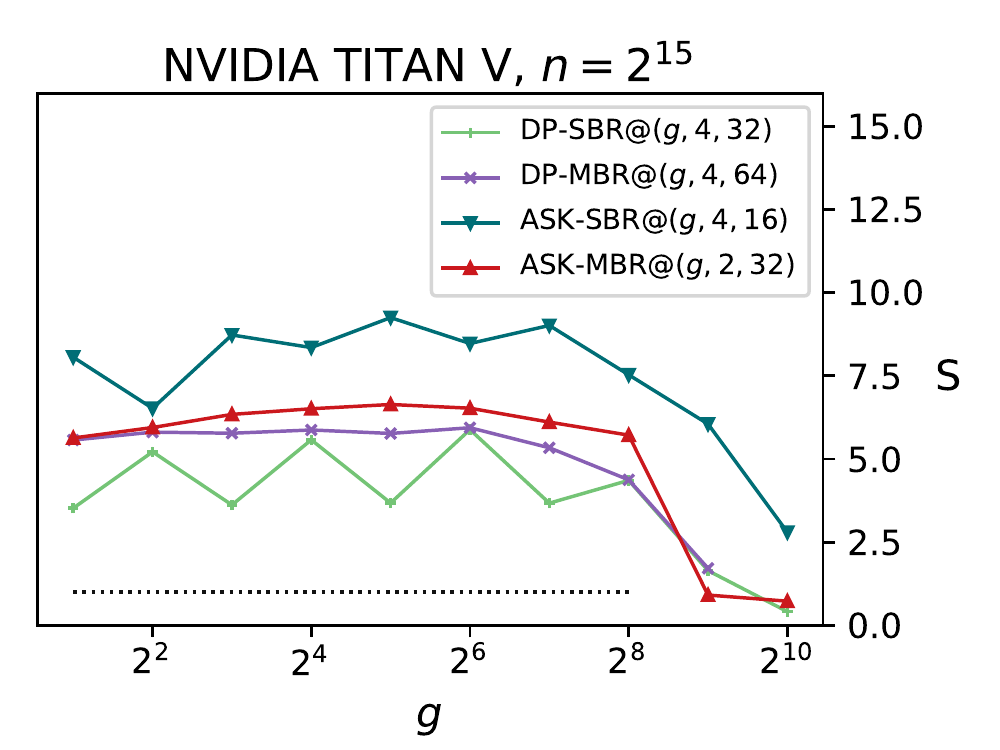}
    \includegraphics[scale=0.44]{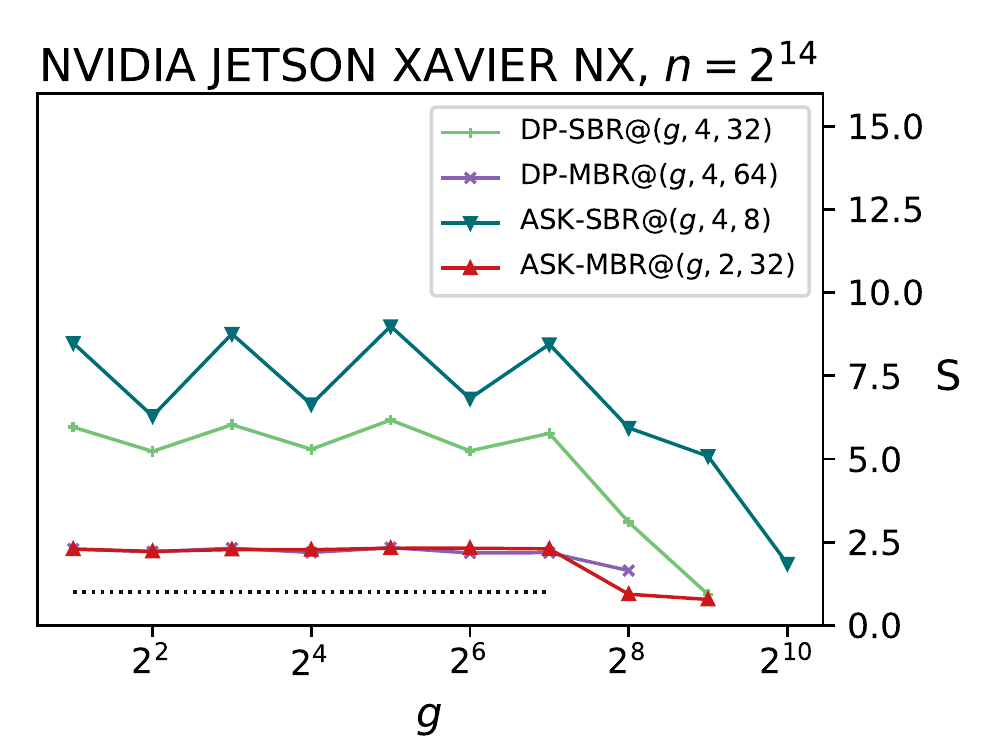}\\
    
    \includegraphics[scale=0.44]{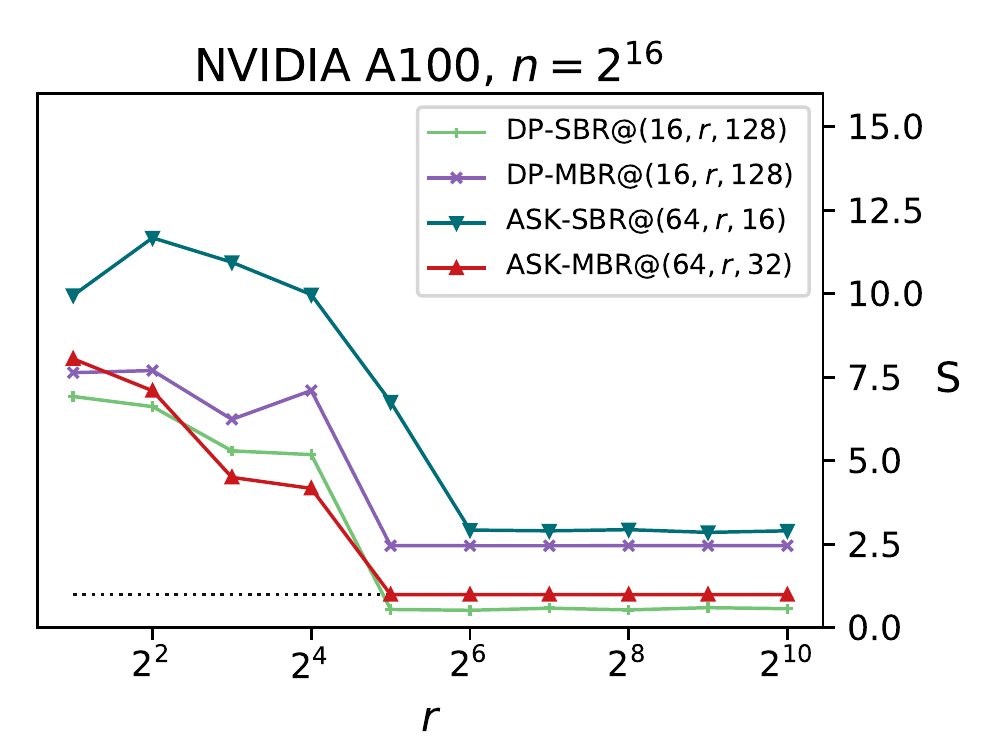}
    \includegraphics[scale=0.44]{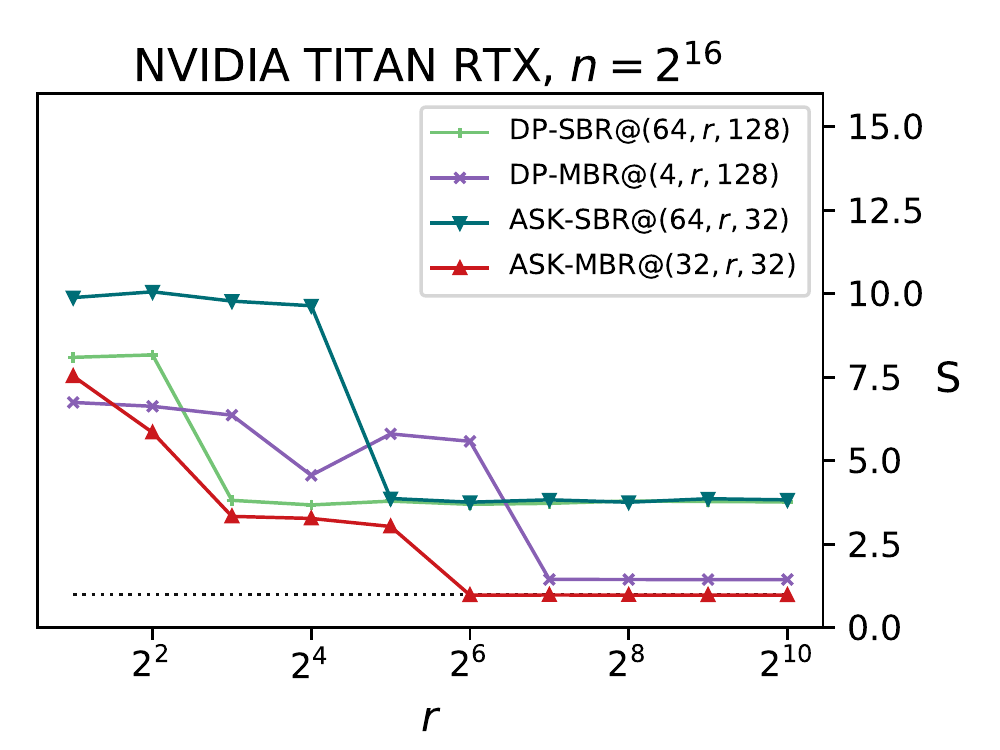}
    \includegraphics[scale=0.44]{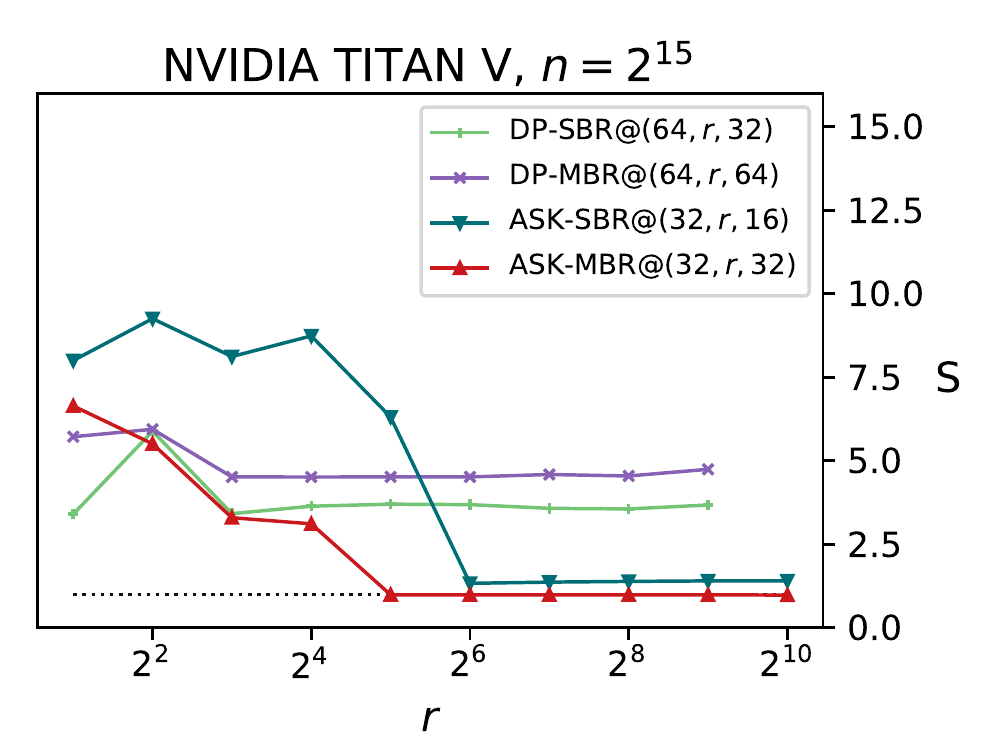}
    \includegraphics[scale=0.44]{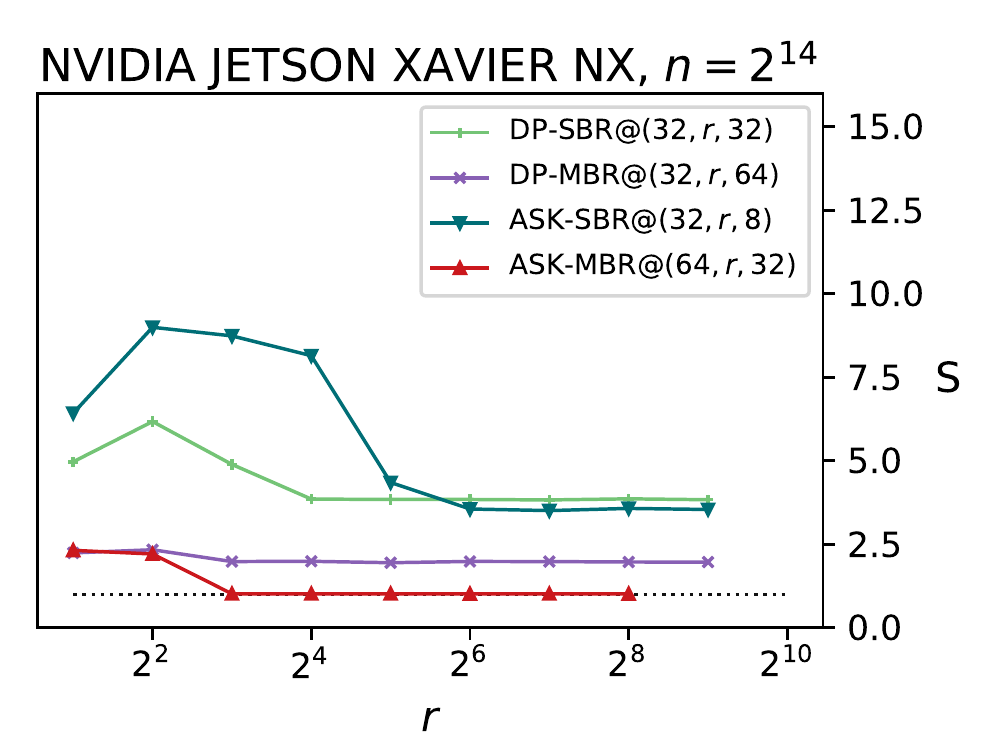}\\
    
    \includegraphics[scale=0.44]{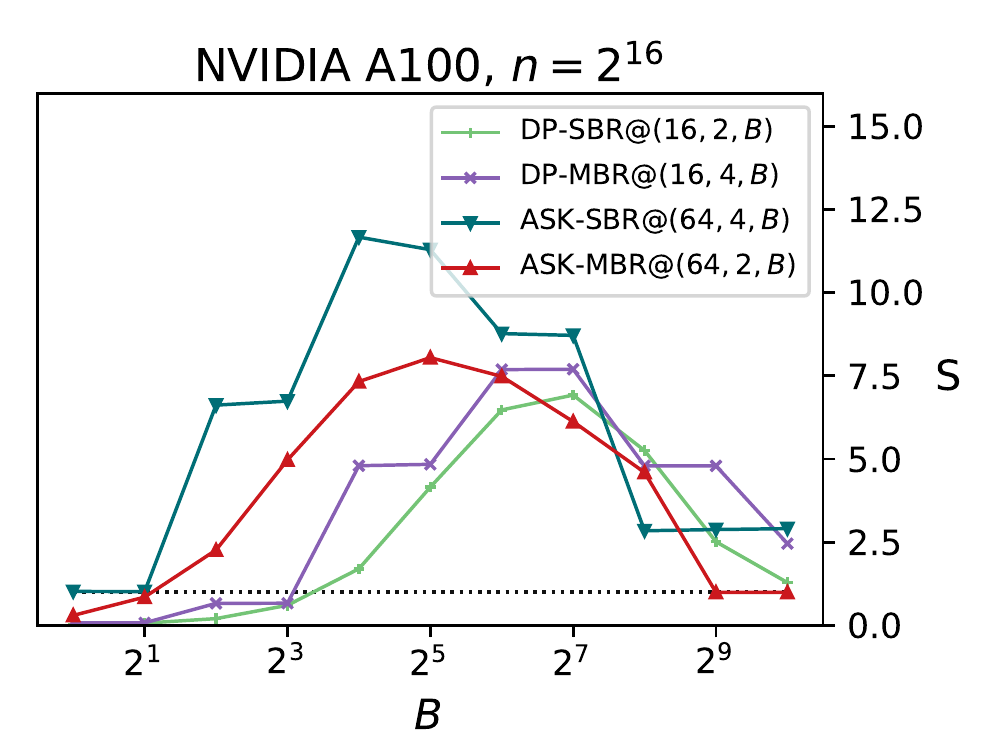}
    \includegraphics[scale=0.44]{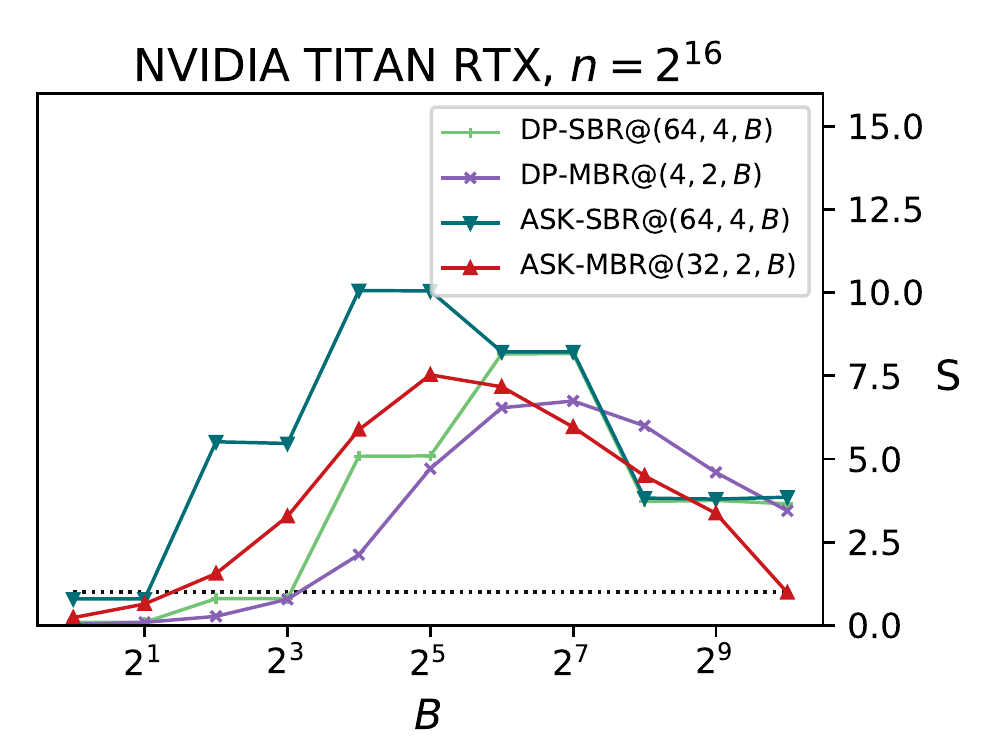}
    \includegraphics[scale=0.44]{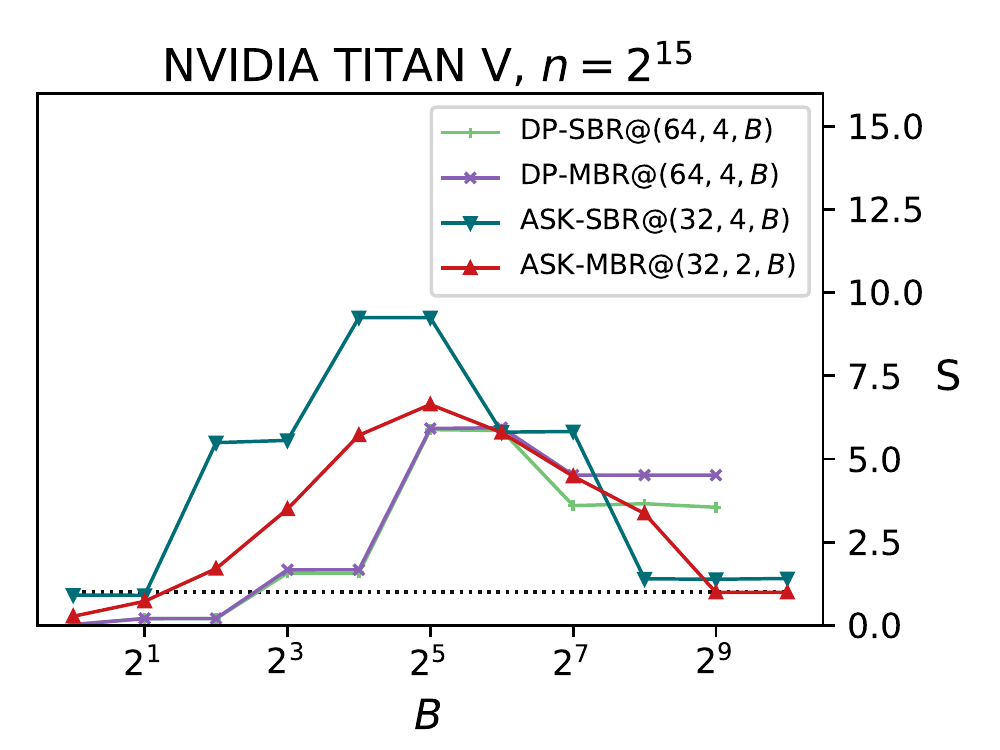}
    \includegraphics[scale=0.44]{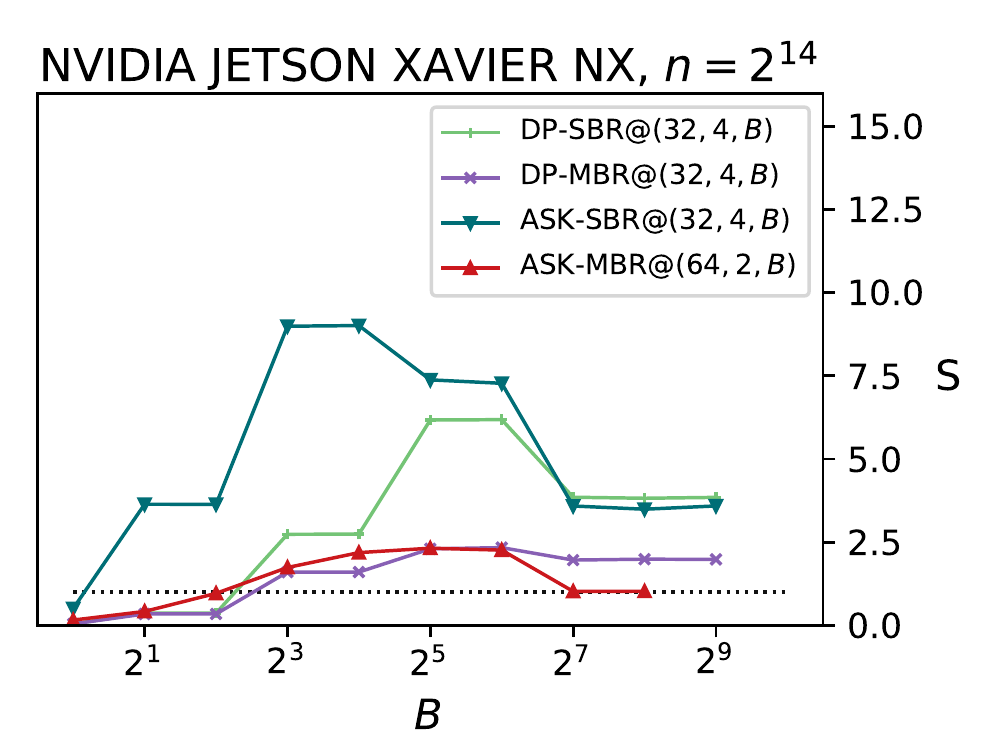}\\
    
    \caption{Speedup curves for DP and ASK approaches with optimal $\{g,r,B\}$ configurations at different problem sizes and its behavior near its optimal configuration.}
    \label{fig:speedup-experimental}
\end{figure*}

Figure \ref{fig:speedup-experimental} presents an experimental set of speedup vs variables $n$, $g$, $r$ and $B$, for the GPUs listed in Table \ref{tab:hardware}. 
On all plots, optimal $\{g,r,B\}$ configurations are used for each chosen combination of GPU, approach and problem size. In the first row of plots, $S$ vs $n$, for all four GPUs the improvements of the DP/ASK variants over the Exhaustive approach start approximately at $n \ge 2^{10}$, with $ASK-SBR$ being the fastest one in all cases, reaching from $9\times$ to $12\times$ of speedup. This speedup take off at $n \ge 2^{10}$ agrees with the theoretical results presented earlier in Figure \ref{fig:speedup-analytic}. The speedup of the rest of the approaches tend to group except in the JETSON XAVIER NX GPU where the MBR variants exhibit a significant reduction in performance.

The second row of plots shows the speedup vs $g$. From the curves one can note that ASK-MBR offers the highest speedup, being maximum at the $g \sim [2^5,2^6]$ region and overall more convenient when $g \le 2^7$. It is worth noticing how ASK-MBR behaves smoother than the other approaches, possibly due to the multiple-block per region scheme being applied at each iteration. This experimental range for $g$ agrees with the theoretical one found in Figure \ref{fig:speedup-analytic}, second row, where the SBR scheme has a non-trivial optimal value in the range $g = [2^5, 2^6]$, while MBR schemes just need $g \le [2^5, 2^7]$ to be satisfied. This experimental behavior also agrees with the theoretical curve of Figure \ref{fig:wrf-analytic}, second row, where large problem sizes make the optimal region for $g$ wider and a less critical decision for the user/programmer. 

The third row of plots shows the speedup vs $r$. For all approaches and for all GPUs, the optimal value is located at the lower values of $r$. This result agrees with the theoretical curve presented in Figure \ref{fig:speedup-analytic}, third row.

The fourth plot presents the speedup vs $B$. For all GPUs the optimal region for $B$ tends to locate at $B \sim 2^5$, which agrees with the theoretical curve presented in Figure \ref{fig:speedup-analytic}.

All experimental plots show that the SBR scheme is faster than MBR, for both DP and ASK. This result differs from the theoretical cost model that puts the MBR scheme with higher speedup than SBR, because of the capability to handle the initial levels of subdivision with more parallel resources. A possible reason for this difference can be that using one block per region (SBR) can favor L1/L2 cache usage and having too many blocks as in MBR may introduce additional scheduling overhead. The current cost model does not consider these aspects. 

When comparing the different GPUs, the highest speedup over the exhaustive approach is achieved with the A100 GPU, followed by the two TITAN GPUs. In the case of the JETSON XAVIER NX GPU the MBR schemes of DP/ASK suffered a significant reduction in performance. This reduction can be related to the small SM count ($q=6$) of that particular chip. In a low SM count scenario, it is highly probable that a multi-block per region (MBR) scheme just introduces additional overhead instead of any performance benefit, as there are not enough SMs to compensate the effort of producing and handling multiple blocks in parallel at each region. Because of this, the single-block per region (SBR) scheme becomes an even more attractive approach for embedded hardware.

\section{Subdivisions at Higher Dimensions}
\label{sec:extending-3D}
When moving to 3$+$ dimensions, additional challenges emerge for both the recursive (DP) and iterative (ASK) subdivision approaches, as the domain is now a $k$-orthotope that generalizes the 2D rectangle. In this Section we describe some of these challenges and provide insights on how they can be handled. 

\subsection{Challenges for DP at 3$+$ Dimensions}
The extension of DP to $k \ge 3$ dimensions would mean that the starting domain is a $k$-orthotope initially subdivided into regions of volume $g^k$, which are $k$-orthotopes as well. Each one of these regions could subdivide into $r^k$ smaller $k$-orthotopes and continue the process recursively analogous to the 2D process. 

A first challenge for DP at $k \ge 3$ dimensions is to be cautious on the total number of recursive kernel calls, as each one introduces a small performance overhead that when added, may constitute a significant part of the running time. Also the number of executing kernels should stay below the maximum number of concurrent kernels that a GPU can handle, other wise the extra kernels will be queued up producing a sequential behavior and potential performance overhead. As of 2022, high-end GPUs can run up to 128 concurrent kernels. One way to handle this challenge is to have low $g,r$ values, and a higher stopping $B$ value, this way the total number of kernel calls is kept up to a margin. 

A second challenge are the limits of GPU compute constructs (grid, block, thread), as they can only be expressed to up to three dimensions, \textit{i.e.}, there are no constructs for $k \ge 4$. Higher dimensional cases can still be implemented in GPUs assuming the programmer builds a higher-dimensional abstraction based on based on 3D constructs. For instance, a 4-dimensional grid can be build by stacking 3D grids which are CUDA-native. In general, a $k$-dimensional space can be represented in terms of $(k-1)$-dimensional ones.  

\subsection{Challenges for ASK at 3$+$ Dimensions}
The extension of ASK to $k \ge 3$ dimensions would involve dealing with sets of $k$-orthotopes at each iteration, identified by the offsets-lookup-table (OLT). A first challenge for ASK would be controlling the OLT's size at each iteration of the subdivision process. An efficient $k$-dimensional OLT scheme can be formulated by inferring from the particular 2D and 3D cases. For the case of two-dimensions, the size of the OLT $T_i^{k=2}$ followed the form
\begin{equation}
    |T_i^{k=2}| = |G_i| \cdot (r_x \times r_y) 
\end{equation}
where all active regions $|G_i|$ of kernel iteration $i$ could subdivide into $r_x \times r_y$ sub-regions. For three-dimensions, sub-regions become $3$-orthotopes and subdivide into $r_x \times r_y \times r_z$ regions, producing an OLT of size
\begin{equation}
    |T_i^{k=3}| = |G_i| \cdot (r_x \times r_y \times r_z).
\end{equation}
For a $k$-dimensional problem, $T_i^k$ would follow the form
\begin{equation}
\label{eq:olt_kdim}
    |T_i^k| = |G_i| \prod_{j=1}^{k} r_j.
\end{equation}
The effect of high-dimensionality affects $|T_i^{k}|$ only at the subdivision factor, \textit{i.e.}, $\prod_{j=1}^k r_i$, and not necessarily through $|G_i|$, because the number of active regions follows a scheme that is specific to the problem itself and not necessarily to the entire embedding grid. This makes it possible to have sizes of $|G_i| \in o(n^k)$ for problems with fractal dimension or PDE simulations on highly sparse domains, among others. In the case of highly dense domains, the upper bound would indeed be $|G_i| \in \Theta(n^k)$, but then the subdivision approach could compensate by doing fewer subdivision levels, or none at all. Also, the $r_j$ values from Eq.~(\ref{eq:olt_kdim}) can be chosen such that $\prod_{j=1}^k r_j \in o(n^k)$; for example $r_j = \log_2(n)$ would produce $\prod_{j=1}^k{\log_2(n)} = \log_2^k(n) < n^k \in o(n^k)$. With this change the OLT's size can be kept asymptotically smaller than the embedding space, $|T_i^k|\ \in o(n^k)$, in both sparse and dense domains.

A second challenge for ASK is the extra space used by storing the $k$-dimensional coordinates explicitly. In this case, an improvement for the OLT at high-dimensions is to store single scalars in the OLT instead of $k$-dimensional coordinates. This compacts the OLT size by an extra factor of $k$. A space filling curve (SFC) can provide the required mapping, \textit{i.e.,}
\begin{align}
    \Omega &: \mathbb{Z}^k \mapsto \mathbb{Z}\\
    \Omega^{-1} &: \mathbb{Z} \mapsto \mathbb{Z}^k.
\end{align}
such that each $k$-dimensional region coordinate is mapped into a unique scalar value through $\Omega$, and recovered back with $\Omega^{-1}$. For the case of two-dimensions ($k=2$), a frequently used SFC is the \textit{canonical order} one which is defined as
\begin{equation}
    \Omega(p)^{k=2} = |G^i|_x \cdot p_y + p_x
\end{equation}
where $|G|_x$ is the hypothetical grid size in the $x$ dimension and $p_x, p_y$ are the given region's top-left corner coordinate as reference. For three-dimensions, $\Omega_{sweep}$ follows the form
\begin{equation}
    \Omega(p)^{k=3} = |G|_y|G|_x p_z + |G|_x p_y + p_x.
\end{equation}
For $k$-dimensions, its general expression would be
\begin{equation}
    \Omega(p)^{k} = \sum_{d=1}^k{\Bigg(p_d \prod_{q=1}^d{|G|_q}}\Bigg)
\end{equation}
Although the canonical order (also known as nested loops) SFC is simple in computation, storing high-dimensional data under this scheme introduces significant cache misses and penalizes performance. Other SFC schemes can be more efficient at handling memory locality at higher dimensions, such as ones based on the Hilbert or Z (also known as Morton order/code) SFCs \cite{bhm2020spacefilling}.

Similar to DP, a third challenge is the native dimensional limit of compute constructs (grid, block, thread). As in DP, higher dimensional cases are still possible to achieve, but will require an adaptation by the programmer, such as stacking 3D dimensional constructs following the SFC scheme selected. 

From all the challenges, the lack of higher-dimensional CUDA constructs is one of the more limiting ones for both DP and ASK, as it requires an advanced programming effort in building an abstraction layer to identify parallel resources (threads and blocks) with $k$-dimensional coordinates.

\section{Discussion and Conclusions}
\label{sec:conclusions}
This work revisited \textit{Dynamic Parallelism} (DP), a GPU programming feature that allows the GPU to employ its parallel resources more efficiently on heterogeneous workloads. A subdivision cost model was formulated to analyze the parameters involved in the cost of the hierarchical subdivision process for problems that exhibit Self Similar Density (SSD) in their heterogeneous data layout. These parameters are the initial grid subdivision ($g$), the exploration subdivision ($r$) and the region size ($B$) where subdivision stops. An experimental study was also conducted using two implementations of the Mandelbrot set; one using the CUDA DP feature (recursive kernels) and the other using a new proposed iterative subdivision scheme based on iterative kernels, named Adaptive Serial Kernels (ASK). Each implementation was evaluated in two schemes, single-block per region (SBR) and multiple-blocks per region (MBR).

The subdivision cost model provided relevant results for the case of heterogeneous problems where the are asymptotically less work-regions than the total embedding space, such as in the computation of the Mandelbrot set. For this case study the model showed that the maximum speedup of a subdivision-based approach is upper bounded by $\mathcal{A}$ which correlates with the dwell of the Mandelbrot process. Also, when doing subdivision, the theoretical cost model suggests that the MBR scheme is potentially faster than the SBR approach, although eventually the two reach the same performance at the large-$n$ regime. In terms of the $\{r,g,B\}$ parameters, the model finds optimal values lie in the range $\{[2,2^6], r \sim 2, B \sim 2^5\}$, values that agreed with the experimental validation.

Experimental results showed that solving the Mandelbrot set with Adaptive Serial Kernels (ASK) as the subdivision approach can be up to $12\times$ faster than the Exhaustive approach, whereas DP is only $7.5\times$ faster. This translates to ASK being up to $60\%$ faster than DP for the same task. Such performance improvement is significant and may compensate the programming effort for many scientific simulations that can take from days to weeks. When comparing the behavior of the different GPUs, we can identify that the high-end ones such as the A100 and TITAN GPUs share a similar behavior and parameter configuration, while the embedded JETSON XAVIER NX showed a significant performance reduction for the MBR schemes of both DP and ASK.  

One aspect that differed between the theoretical and experimental results is that the theoretical model suggested that MBR schemes would be more efficient than SBR ones, however experimentation proved the contrary; SBR was faster in all cases. An explanation for this can be the fact that there are some overheads in the GPU thread-block scheduling process that can make MBR work slower in practice, as the model does not consider these overheads. Also, the SBR approach could favor L1/L2 hit ratios more than in MBR, as it reuses the same thread-block for the entire region.

Future work can focus on improving the cost model to consider the GPU overhead introduced by the MBR approach, which ended up being slower than SBR in empirical tests. Another future research can be to explore if Tensor and Ray-Tracing cores could further improve the efficiency of a subdivision-based process. Recent works have shown that Tensor cores can be used to accelerate several workloads different from AI, including thread mapping, arithmetic reductions and prefix-sums \cite{QUEZADA202210, NAVARRO2020158, navarro2020gpu, dakkak2019accelerating}. Expressing the subdivision arithmetic as Matrix Multiply Accumulate (MMA) operations could further improve performance assuming that the benefits overcome the cost of the extra data movement required from memory to the matrix registers (fragments in CUDA). On the other hand, the use of Ray-Tracing (RT) cores allows exploring unstructured space efficiently in GPU and recent works have also shown that using them on non-graphical applications can still provide significant performance improvement \cite{zellmann2020accelerating, salmon2019exploiting, evangelou2021fast}. The use of RT-cores could help in rapidly detecting the location of dense workloads in an heterogeneous data-parallel domain mapped into space.  

\section*{Acknowledgement}
This work was supported by FONDECYT grant \#1221357, the Temporal research group and the Patagón supercomputer of Universidad Austral de Chile (FONDEQUIP EQM180042).
\bibliographystyle{elsarticle-num}
\bibliography{main}
\end{document}